\documentclass[preprint,onecolumn,nofootinbib]{revtex4}
\pdfoutput=1
\usepackage[colorlinks=true,linkcolor=blue,urlcolor=blue,filecolor=black,citecolor=red,pdfstartview=FitV,pdftitle={},pdfsubject={},pdfkeywords={},pdfpagemode=None,bookmarksopen=true]{hyperref}
\usepackage{graphicx}
\usepackage{amsmath}
\usepackage{amsfonts}
\usepackage{amssymb,ulem}
\usepackage{color}%
\usepackage{dcolumn}
\newcommand{\f}{\begin{equation}}
\newcommand{\ff}{\end{equation}}
\newcommand{\fa}{\begin{eqnarray}}
\newcommand{\ffa}{\end{eqnarray}}
\newcommand{\bsub}{\begin{subequations}}
\newcommand{\esub}{\end{subequations}}

\begin{document}
\title{Holographic p-wave superconductivity from higher derivative theory}
\author{Yan Liu$^{1}$}
\author{Guoyang Fu$^{2}$}
\author{Hai-Li Li$^{1}$}
\author{Jian-Pin Wu$^{2,3}$}
\thanks{jianpinwu@yzu.edu.cn}
\author{Xin Zhang$^{1}$}
\thanks{zhangxin@mail.neu.edu.cn}
\affiliation{
$^1$ College of Sciences \& MOE Key Laboratory of Data Analytics and Optimization for Smart Industry, Northeastern University, Shenyang 110819, China\ \\
$^2$ Center for Gravitation and Cosmology, College of Physical Science and Technology, Yangzhou University, Yangzhou 225009, China\ \\
$^{3}$ School of Aeronautics and Astronautics, Shanghai Jiao Tong University, Shanghai 200240, China}
\begin{abstract}
We construct a holographic SU(2) p-wave superconductor model with Weyl corrections. The high derivative (HD) terms do not seem to spoil the generation of the p-wave superconducting phase. We mainly study the properties of AC conductivity, which is absent in holographic SU(2) p-wave superconductor with Weyl corrections. The conductivities in superconducting phase exhibit obvious anisotropic behaviors. Along $y$ direction, the conductivity $\sigma_{yy}$ is similar to that of holographic s-wave superconductor. The superconducting energy gap exhibits a wide extension. For the conductivity $\sigma_{xx}$ along $x$ direction, the behaviors of the real part in the normal state are closely similar to that of $\sigma_{yy}$. However, the anisotropy of the conductivity obviously shows up in the superconducting phase. A Drude-like peak at low frequency emerges in $Re\sigma_{xx}$ once the system enters into the superconducting phase, regardless of the behaviors in normal state.
\end{abstract}
\maketitle
\section{Introduction}
The AdS/CFT correspondence \cite{Maldacena:1997re,Gubser:1998bc,Witten:1998qj,Aharony:1999ti}, linking a gravity theory in a $(d+1)$ dimensional AdS spacetime to a conformal field theory on its $d$ dimensional boundary, has provided very valuable insights into the condensed matter theory (CMT). The holographic superconductor is one of the most successful applications of AdS/CFT in CMT. Some universal properties of high $T_{c}$ superconductor are addressed in holographic superconductor \cite{Hartnoll:2008vx,Hartnoll:2008kx,Horowitz:2010gk}. Especially, the superconducting energy gap $\omega_g/T_c$ is approximately $8$, which belongs to the region of some high $T_{c}$ superconductor materials~\cite{Gomes:2007}.

The simple holographic superconductor model \cite{Hartnoll:2008vx,Hartnoll:2008kx,Horowitz:2010gk} is built in the large N limit. It is interesting and important to implement the effect beyond the large N limit in holographic superconductor model. The top-down embedding of these holographic superconductors in string theory can be found in \cite{Gauntlett:2009dn,Gauntlett:2009bh}. However, before the string theory is completely understood, an alternative scheme is to include the higher curvature/derivative corrections to the gravity, which provides an effective method to address these problems. It is because the higher curvature/derivative corrections on the gravity side lead to the finite coupling correction on the boundary field theory, which provides us a large class of holographic effective field theories we can study. Based on this idea, several holographic superconductor models have been implemented in Gauss-Bonnet (GB) gravity framework in which a high curvature gravity theory contains only the curvature-squared interaction (see \cite{GBHS1,GBHS2,GBHS3,GBHS4,GBHS5,GBHS6,GBHS7} and references therein). It is found that the superconducting energy gap in the holographic superconductor model with GB correction is larger than that of the standard version of holographic superconductor. This observation is further confirmed in the holographic superconductor model built in the framework of quasi-topological gravity, which contains both the curvature-squared interaction and the curvature-cubed interaction \cite{HSinQT1,HSinQT2,Kuang:2011dy}.

Another important high derivative (HD) term is the Weyl tensor coupled with the Maxwell field. The AC (alternating current) from this HD term on the Schwarzschild-AdS (SS-AdS) is deeply studied in \cite{Myers:2010pk,Sachdev:2011wg,Hartnoll:2016apf,Ritz:2008kh,WitczakKrempa:2012gn,
WitczakKrempa:2013ht,Witczak-Krempa:2013nua,Witczak-Krempa:2013aea,Katz:2014rla}. In $4$ derivative theory \cite{Myers:2010pk,Sachdev:2011wg,Hartnoll:2016apf,Ritz:2008kh,WitczakKrempa:2012gn,
WitczakKrempa:2013ht,Witczak-Krempa:2013nua}, a so-called Damle-Sachdev (DS) peak resembling the particle response \cite{Damle:1997rxu}, or a dip being similar to the vortex response emerge at the low frequency depending on the sign of the coupling parameter. These behaviors are analogous to that of the superfluid-insulator quantum critical point \cite{Myers:2010pk,Sachdev:2011wg,Hartnoll:2016apf}. Further, when $6$ derivative terms are considered, depending on the sign of coupling parameter, a sharp Drude-like peak or a hard-gap-like are exhibited at low frequency \cite{Witczak-Krempa:2013aea}. Further, based on the HD theory from Weyl tensor coupled Maxwell field, we have constructed the holographic superconductor model \cite{Wu:2010vr,Wu:2017xki}. An important characteristic is that the running of the superconducting energy gap ranging from $5.5$ to $16.2$. It suggests that in this holographic model, one can model the weakly coupled BCS theory and also model some high $T_c$ superconductor materials. Then, lots of holographic superconductor models in this HD framwork are constructed; see \cite{Ling:2016lis,Huang:2019yov,Momeni:2014efa,Momeni:2012uc,Momeni:2011ca,Ma:2011zze,Momeni:2013fma,Mansoori:2016zbp} and references therein.

In this paper, we are interested in the effects of Weyl corrections on the p-wave superconductors. A holographic p-wave superconductor model was first proposed by Gubser and Pufu \cite{Gubser:2008wv}. They introduced a SU(2) Yang-Mill gauge field, which leads to the superconducting phase transition. The conductivities of this model exhibited remarkable anisotropic behaviors. Then, lots of holographic p-wave superconductor models are proposed, see \cite{Zayas:2011dw,Ammon:2008fc,Basu:2008bh,Ammon:2009fe,Cai:2013pda} and references therein. One of them is the so-called holographic MCV (Maxwell complex vector) p-wave superconductor \cite{Cai:2013pda}. In this model the vector condensate is induced by a magnetic field, meaning that the emergence of phase transition is induced not only by the chemical potential/charge density, but also by the background magnetic field. Based MCV p-wave model, the effects of the Weyl corrections is deeply studied in~\cite{Huang:2019yov,Lu:2020phn}. It is found that the presence of the higher order Weyl corrections facilitates the emergence of the superfluid phase~\cite{Huang:2019yov}. In the real part of conductivity a Drude-like peak or an obviously pronounced peak emerges at the low frequency and at the intermediate frequency an energy gap emerges when temperature is enough low, which depend on the coupling parameters~\cite{Lu:2020phn}. Similarly, in the SU(2) p-wave model with Weyl corrections, it is found that the condensation becomes harder with the increase of the coupling parameter~\cite{Momeni:2012ab}. However, the frequency dependent conductivity in SU(2) p-wave model with Weyl corrections is still absent so far. Therefore, in this work we will investigate the effect on holographic SU(2) p-wave superconducting phase transition when the higher order Wely corrections are considered. Further, we also calculate the frequency dependent conductivity and make an analysis.

Our work is organized as follows. In Sect.~\ref{sec-setup}, we construct the holographic $p$-wave superconductor with Weyl corrections.  The numerical results for the condensation of the $p$-wave superconductor are presented in Sect.~\ref{sec-con}. Then we numerically calculate the conductivity  in Sect.~\ref{sec-conductivity}. Conclusions follow in Sect.~\ref{sec-Conclusions}.

\section{Holographic framework}\label{sec-setup}

We start with the bulk theory of the Einstein gravity including the coupling between Weyl tensor and SU(2) Yang-Mills field in a $4$-dimensional spacetime (EYM-Weyl model), which action reads as
\fa
\label{action}
S=\int d^4x\sqrt{-g}\Big[\frac{1}{2\kappa^2}\Big(R+\frac{6}{L^2}\Big)
-\frac{L^2}{8g_F^2}(F_{\mu\nu})_aX^{\mu\nu\rho\sigma}(F_{\rho\sigma})^a\Big]
\,,
\ffa
where $\kappa$, $g_F$ and $L$ are the gravitational constant, Yang-Mills coupling constant and the AdS radius, respectively.
The Yang-Mills field strength is
\fa
\label{YMs}
(F_{\mu\nu})^a=\partial_{\mu}A_{\nu}^a-\partial_{\nu}A_{\mu}^a+\epsilon^a_{\ bc}A_{\mu}^bA_{\nu}^c\,,
\ffa
where $\epsilon^{abc}$ is the totally antisymmetric tensor. $X$ is an infinite family of HD terms as \cite{Witczak-Krempa:2013aea}
\fa
X_{\mu\nu}^{\ \ \rho\sigma}&=&
I_{\mu\nu}^{\ \ \rho\sigma}-8\gamma_{1,1}L^2 C_{\mu\nu}^{\ \ \rho\sigma}
-4L^4\gamma_{2,1}C^2I_{\mu\nu}^{\ \ \rho\sigma}
-8L^4\gamma_{2,2}C_{\mu\nu}^{\ \ \alpha\beta}C_{\alpha\beta}^{\ \ \rho\sigma}
\nonumber
\\
&&
-4L^6\gamma_{3,1}C^3I_{\mu\nu}^{\ \ \rho\sigma}
-8L^6\gamma_{3,2}C^2C_{\mu\nu}^{\ \ \rho\sigma}
-8L^6\gamma_{3,3}C_{\mu\nu}^{\ \ \alpha_1\beta_1}C_{\alpha_1\beta_1}^{\ \ \ \alpha_2\beta_2}C_{\alpha_2\beta_2}^{\ \ \ \rho\sigma}
+\ldots
\,,
\label{X-tensor}
\ffa
where $I_{\mu\nu}^{\ \ \rho\sigma}=\delta_{\mu}^{\ \rho}\delta_{\nu}^{\ \sigma}-\delta_{\mu}^{\ \sigma}\delta_{\nu}^{\ \rho}$
is an identity matrix and $C^n=C_{\mu\nu}^{\ \ \alpha_1\beta_1}C_{\alpha_1\beta_1}^{\ \ \ \alpha_2\beta_2}\ldots C_{\alpha_{n-1}\beta_{n-1}}^{\ \ \ \mu\nu}$.
$L$ is introduced in the above equations such that the coupling parameters $g_F$ and $\gamma_{i,j}$ are dimensionless. In this paper, we shall truncate the action up to the $6$ derivative terms. Since the effect of both the $6$ derivative terms are similar, we shall set $\gamma_{2,2}=0$.
In addition, for later convenience, we denote $\gamma_{1,1}=\gamma$ and $\gamma_{2,1}=\gamma_1$.

When HD terms \eqref{X-tensor} are included, the equations of motion (EOMs) for the above system become a set of differential equations beyond second order and with high nonlinearity, which is a hard task to solve this system with full backreaction. However, one can capture some qualitative properties in the so-called probe limit, where the back reaction of Yang-Miles field on the background is ignored. Indeed, from the action \eqref{action}, it is easy to find that if we let $\kappa^2/g_F^2\ll 1$, we can safely ignore the back reaction of Yang-Miles field and Weyl tensor on the background.
In this case, the background geometry is just the SS-AdS black brane
\fa
\label{bl-br}
&&
ds^2=\frac{L^2}{u^2}\Big(-f(u)dt^2+dx^2+dy^2\Big)+\frac{L^2}{u^2f(u)}du^2\,,
\nonumber
\\
&&
f(u)=(1-u)p(u)\,,~~~~~~~
p(u)=u^2+u+1\,.
\ffa
$u=0$ is the asymptotically AdS boundary while the horizon locates at $u=1$. The Hawking temperature of this system is $\hat{T}=3/4\pi L^2$. Then, it is straightforward to derive the Yang-Mills EOM from the action \eqref{action}, which reads as
\fa
&&
\nabla_{\mu}(F^{a\mu\nu}-4\gamma L^2C^{\mu\nu\rho\sigma}-4L^4\gamma_{1}C^2F^{a\mu\nu})
\nonumber
\\
&&
=-\epsilon^a_{bc}{A^b_\mu}F^{c\mu\nu}+4\gamma L^2C^{\mu\nu\rho\sigma}\epsilon^a_{bc}{A^b_\mu}F^c_{\rho\sigma}+4\gamma_{1}C^2\epsilon^a_{bc}{A^b_\mu}F^{c\mu\nu}\,.
\label{EOM-YM-o}
\ffa
For the Maxwell-Weyl system with a coupling between U(1) gauge field and Weyl tensor studied in \cite{Ritz:2008kh,Myers:2010pk,Witczak-Krempa:2013aea}, when the other parameters are turned off, $\gamma$ and $\gamma_1$ are confined to the ranges $-1/12\leq\gamma\leq1/12$ \cite{Ritz:2008kh,Myers:2010pk} and $\gamma_1\leq 1/48$ \cite{Witczak-Krempa:2013aea} on top of SS-AdS black brane, respectively. These constraints originate from the instabilities and causality of the vector modes. However, for EYM-Weyl model, the instabilities and causality of the vector modes must be reexamined and analyzed, which we leave for future. In this paper, we shall constraint the coupling parameters $\gamma$ and $\gamma_1$ in small space of parameter, which is in general safe. In what follows, we shall set $L=1$ and $g_F=1$ for convenience.

\section{Condensation}\label{sec-con}

In this section, we are going to investigate the holographic p-wave superconducting phase transition with HD terms. To this end, following the strategy presented in \cite{Gubser:2008wv}, we take the following ansatz
\fa
\label{ansatz}
A=\phi(u)\tau^3dt+\psi(u)\tau^1dx\,,
\ffa
where $\tau^i$ with $i=1,2,3$ are the $SU(2)$ generators satisfying the commutation relation $[\tau^i,\tau^j]=\epsilon^{ijk}\tau^k$.
The nonzero $\psi(u)$ breaks the U(1) gauge symmetry generated by $\tau^3$ and results in the superconducting phase transition in the dual boundary field theory. Since we choose x-axis as the special direction, the operator $<J_x^1>$ in the boundary field theory, which is dual to $\psi(u)$ in bulk, breaks the rotational symmetry and so we interpret it as a p-wave superconducting phase transition.

Under the above ansatz, we can explicitly write down the Yang-Mills equations
\fa
&&
f^2(4u^4{\gamma_1}{f''}^2+2u^2{\gamma}f''-3)\psi''+(f{f'}(4u^4{\gamma_1}{f''}^2+2u^2\gamma{f''}-3)\psi'
\nonumber
\\
&&
+(4u^4{\gamma_1}{f''}^2+2u^2\gamma{f''}-3)\phi^2\psi+2uf^2(\gamma+4u^2{\gamma_1}f'')(2f''+uf'''))=0\,,
\label{YM-equation1}
\\
&&
f(-4u^4{\gamma_1}{f''}^2+4u^2\gamma{f''}+3)\phi''+4uf(\gamma-2u^2{\gamma_1}f'')(2f''+uf''')\phi'
\nonumber
\\
&&
+(4u^4{\gamma_1}{f''}^2+2u^2\gamma{f''}-3)\phi\psi^2=0\,.
\label{YM-equation2}
\ffa
Irrespective of any details of the above EOMs, we have the asymptotical behaviors of $\psi$ and $\phi$ at the conformal boundary as
\fa
&&
\psi=\psi^{(0)}+\psi^{(1)}u\,,
\label{boundary1}
\\
&&
\phi=\mu-{\rho}u\,,
\label{boundary2}
\ffa
where $\mu$ and $\rho$ are the chemical potential and charge density of the dual field theory, respectively. We take the standard quantization, for which $\psi^{(0)}$ is the source and $\psi^{(1)}$ stands for the expectation value of the $<J_x^1>$ operator. And then, we set source $\psi^{(0)}=0$, which make sure that the condensate is not sourced. Now, this system is determined by a dimensionless quantity $T\equiv\hat{T}/\mu$, as well as the model parameters, i.e., $\gamma$ and $\gamma_1$.

\begin{figure}
\center{
\includegraphics[scale=0.6]{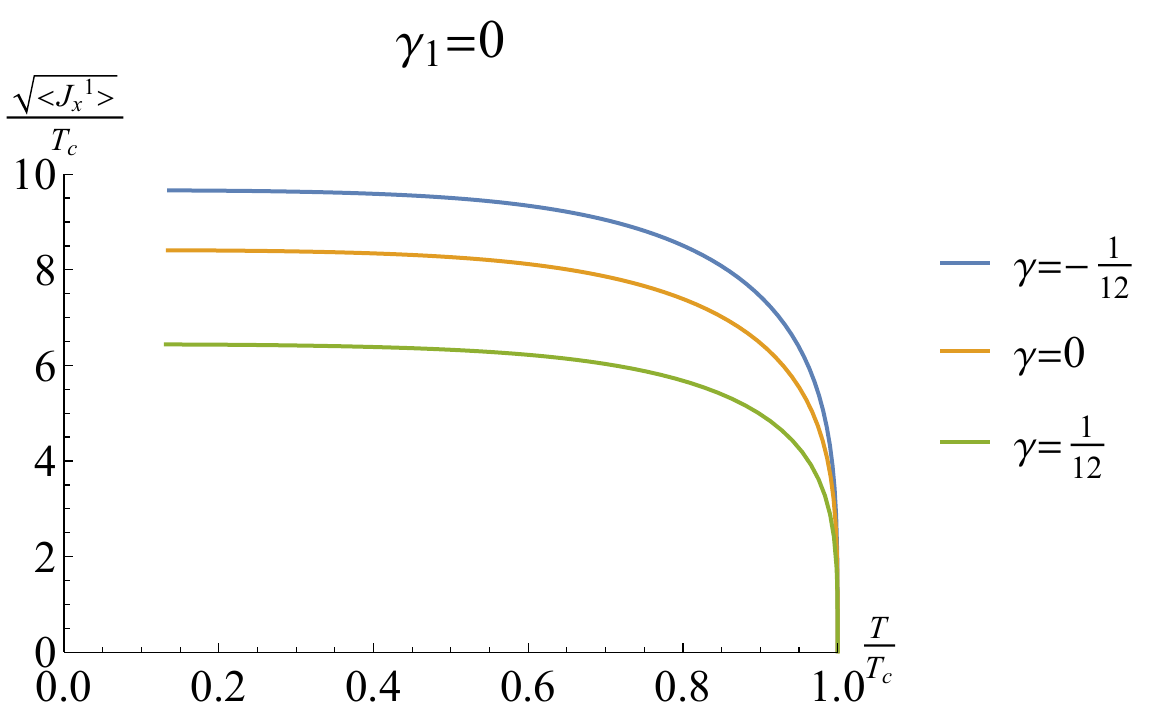}\ \hspace{0.5cm}
\includegraphics[scale=0.6]{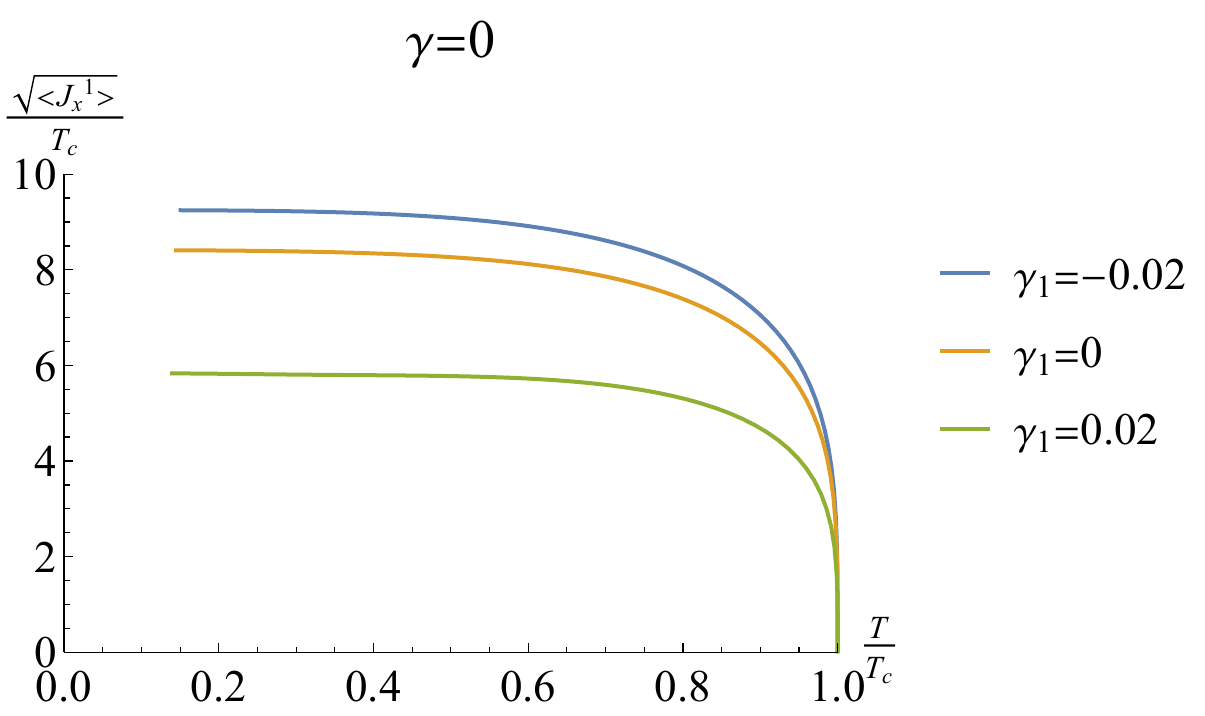}\ \\
\caption{\label{fig-Jx1} The condensation $<J_x^1>$ as a function of temperature. Left plot is for the $4$ derivative theory and the right plot is for the $6$ derivative one.}}
\end{figure}
At the horizon, it is required that both fields $\psi(u)$ and $\phi(u)$ are regular. Combing it with the boundary conditions~(\ref{boundary1}) and~(\ref{boundary2}) at the boundary, we can numerically solve
the coupling EOMs (\ref{YM-equation1}) and (\ref{YM-equation2}) by using a shooting method and study the condensation behavior. The results are exhibited in FIG.\ref{fig-Jx1} to demonstrate the condensation $\sqrt{<J_x^1>}/T_{c}$ as a function of the temperature $T/T_{c}$. The figures explicitly describe that when we cool the system such that the temperature is lower than some critical temperature $T_c$, the superconducting phase transition happens. It indicates that the holographic p-wave superconducting phase transition can develop in the higher derivative theory as that of s-wave superconductor studied in \cite{Wu:2010vr,Wu:2017xki}. As the temperature is further tuned lower, the condensation $\sqrt{<J_x^1>}$ goes to a constant, which indicates that the condensation phase becomes stable. In particular, we observe that as the coupling parameter $\gamma$ or $\gamma_1$ is tuned smaller, the stable condensation value increases. It implies a larger superconducting energy gap $\omega_g/T_c$ in the frequency dependent conductivity along $y$ direction, which shall be explicitly demonstrated below. These observations are qualitatively similar to that of the holographic s-wave superconductor from higher derivative theory~\cite{Wu:2010vr,Li:2019dmm,Wu:2017xki}.

Also we work out the critical temperature $T_c$ of superconducting phase transition for some model parameters, which are shown in TABLE \ref{Tcgamma} and TABLE \ref{Tcgamma1}. From these table, we see that with the decrease of the model parameters, $T_c$ goes down. It suggests that the condensation becomes harder when the model parameters $\gamma$ or $\gamma_1$ decrease. This result is also consistent with that of s-wave superconductor from higher derivative theory~\cite{Wu:2010vr,Li:2019dmm,Wu:2017xki}.

\begin{widetext}
\begin{table}[ht]
\begin{center}

\begin{tabular}{|c|c|c|c|c|c|c|c}
         \hline
$~~\gamma~~$ &~~$-1/12$~~&~~$0$~~&~~$1/12$~~
          \\
        \hline
~~$T_{c}$~~ & ~~$0.0569$~~ & ~~$0.0654$~~ & ~~$0.0854$~~
          \\
        \hline
\end{tabular}
\caption{\label{Tcgamma} The critical temperature $T_{c}$ for different coupling parameter $\gamma$. Here $\gamma_{1}=0$.}
\end{center}
\end{table}
\end{widetext}

\begin{widetext}
\begin{table}[ht]
\begin{center}

\begin{tabular}{|c|c|c|c|c|c|c|c}
 \hline
$~~\gamma_{1}~~$ &~~$-0.02$~~&~~$0$~~&~~$0.02$~~
          \\
        \hline
~~$T_{c}$~~ & ~~$0.0595$~~ & ~~$0.0654$~~ & ~~$0.0942$~~
          \\
        \hline
\end{tabular}
\caption{\label{Tcgamma1} The critical temperature $T_{c}$ for different coupling parameter $\gamma_{1}$. Here $\gamma=0$.}
\end{center}
\end{table}
\end{widetext}

\section{superconductivity}\label{sec-conductivity}

Lots of works have been developed to build holographic p-wave superconductor models with Weyl corrections and the properties of superconducting phase transition are also studied~\cite{Huang:2019yov,Zhang:2015eea,Lu:2020phn,Zhao:2012kp,Momeni:2012ab}. However, as far as we know, the frequency dependent conductivity~(alternating current conductivity, AC conductivity) of holographic p-wave superconductor with Weyl corrections is still absent. In this section, we shall concentrate on this topic.

To study the AC conductivity of the system, we turn on the following consistent linear perturbation of the $SU(2)$ gauge field
\fa
\label{perturbation}
\delta A(t,u)= e^{-iwt}[(A^{1}_{t}(u)\tau^{1}+A^{2}_{t}(u)\tau^{2})dt+A^{3}_{x}(u)\tau^{3}dx+A^{3}_{y}(u)\tau^{3}dy]\,.
\ffa
Above we have assumed that the perturbation has a time dependent form as $e^{-iwt}$. Plugging the above perturbation into Yang-Mill EOM \eqref{EOM-YM-o}, we have four ordinary differential equations for  $A^3_y$, $A^3_x$, $A^1_t$ and $A^2_t$. It is easy to find that the equation of $A^3_y$ decouples from that of other components: $A^3_x$, $A^1_t$ and $A^2_t$, which couple together. It suggests that the conductivity of holographic p-wave superconductor exhibits anisotropy, which is the characteristic different from the s-wave superconductor. Next, we study the characteristics of the conductivities $\sigma_{yy}$ and $\sigma_{xx}$. In particular, we shall mainly concentrate on the effects from the Weyl corrections.

\subsection{$\sigma_{yy}$}
To calculate the conductivity $\sigma_{yy}$, we only need the EOM for $A_{y}^{3}$, which reads as
\fa
{A_{y}^{3}}''+\Big(\frac{f'}{f}+\frac{2u(\gamma+4u^{2}{\gamma_{1}}f'')(2f''+uf''')}{-3+2u^{2}f''({\gamma}+2u^{2}{\gamma_{1}}f'')}\Big){A_{y}^{3}}'+
\nonumber
\\
\Big(\frac{\omega^{2}}{f^2}+\frac{\psi^{2}}{f}\Big)\Big(\frac{6u^{2}\gamma f''}{2u^{2}f''(2u^{2}\gamma_{1}f''+\gamma)-3}-1\Big)
A_{y}^{3}=0\,.
\label{Ay-eom}
\ffa
This perturbative equation is similar to that of the holographic s-wave superconductor studied in \cite{Wu:2010vr,Wu:2017xki}. At infinity boundary ($z\rightarrow 0$), the perturbation $A_{y}^{3}$ falls in the following form
\fa
\label{Ay-behavior-inf}
{A_{y}^{3}(u)=A_{y}^{3(0)}+A_{y}^{3(1)}u}\,,
\ffa
According to the holographic dictionary, the conductivity reads as
\fa
\label{sigma-def}
\sigma(\omega)=-\frac{i}{\omega}\frac{A^{3(1)}_{y}}{A^{3(0)}_{y}}\,.
\ffa
Numerically solving the perturbative equation \eqref{Ay-eom} with ingoing boundary condition at the horizon, we can read off $A^{3(0)}_{y}$ and $A^{3(1)}_{y}$.

\begin{figure*}
\center{
\includegraphics[scale=0.6]{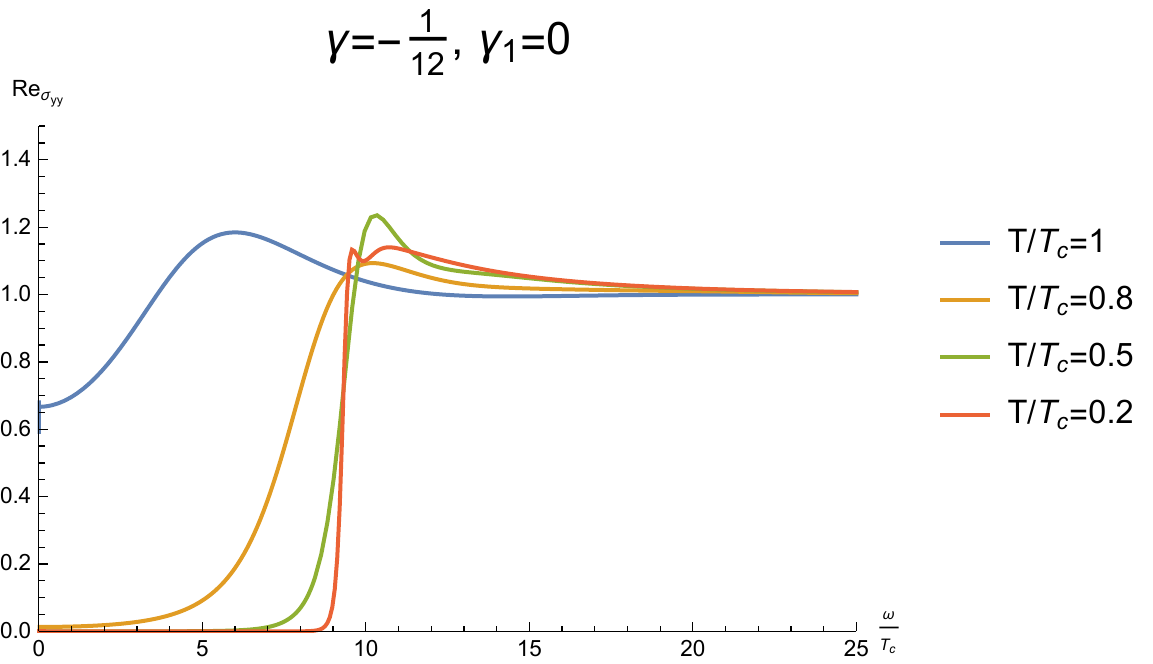}\hspace{0.5cm}
\includegraphics[scale=0.6]{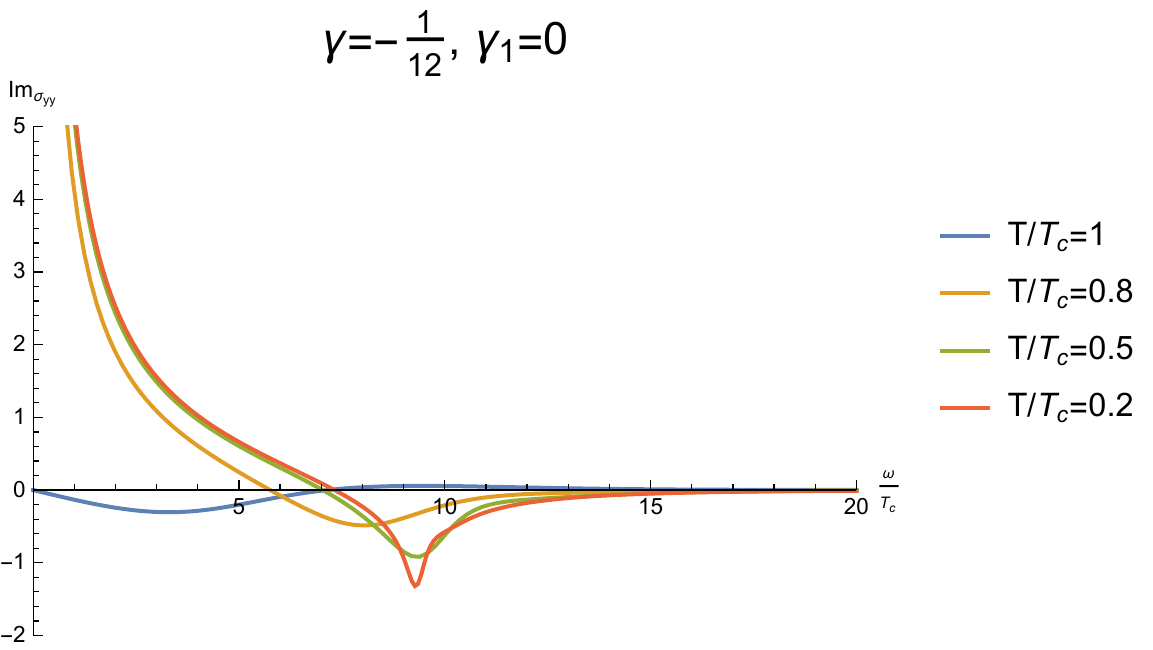} \\
\includegraphics[scale=0.6]{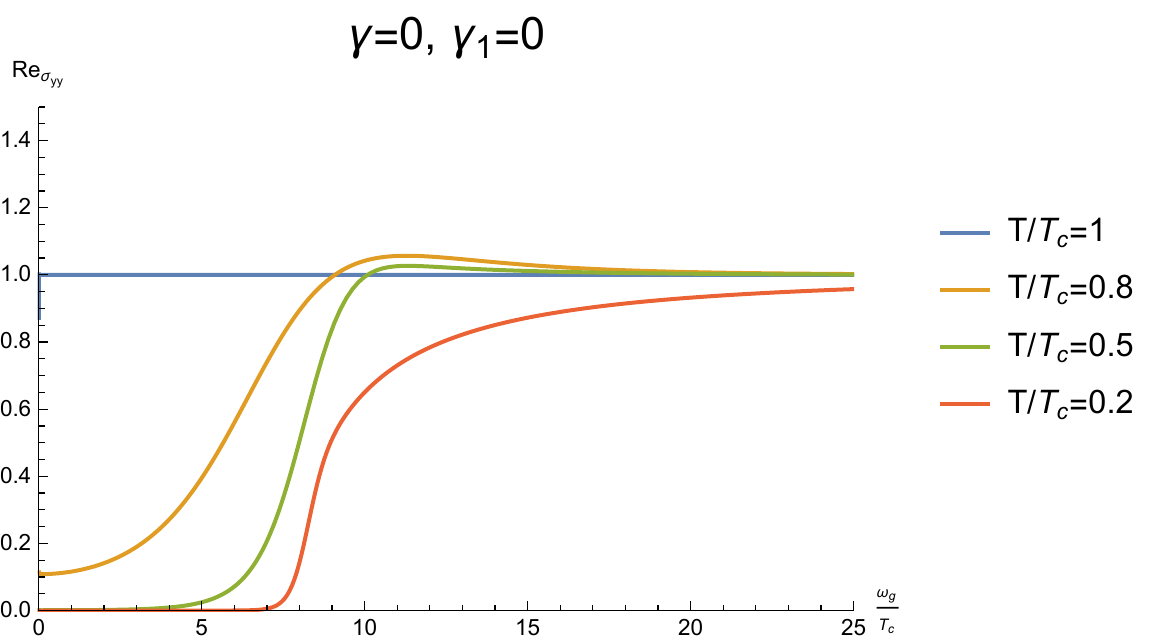}\hspace{0.5cm}
\includegraphics[scale=0.6]{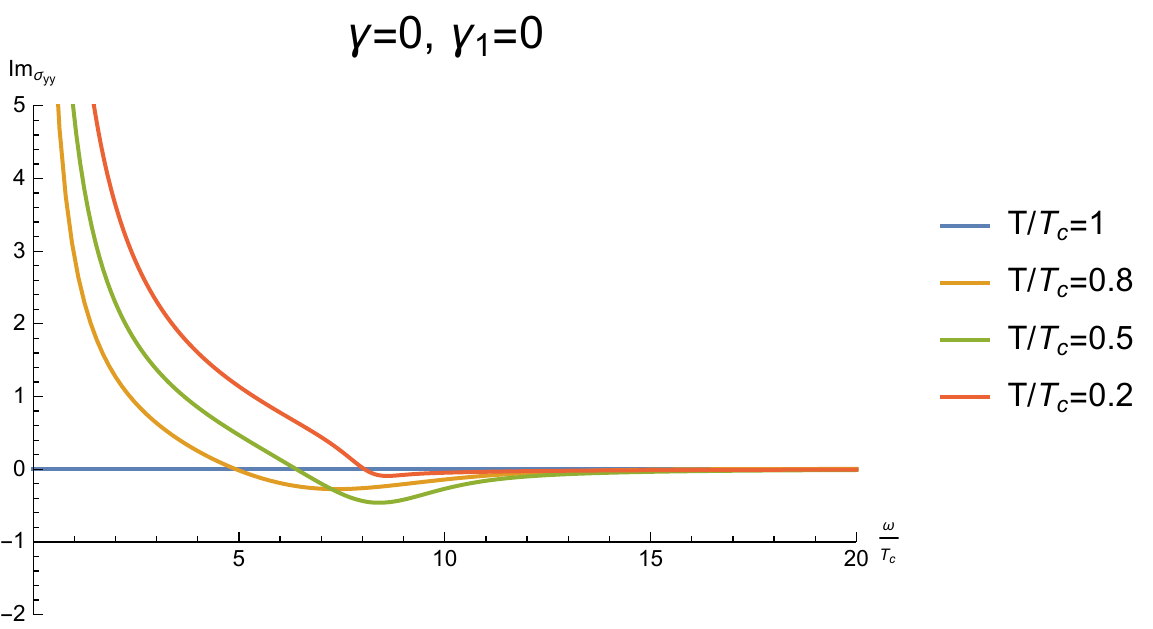} \\
\includegraphics[scale=0.6]{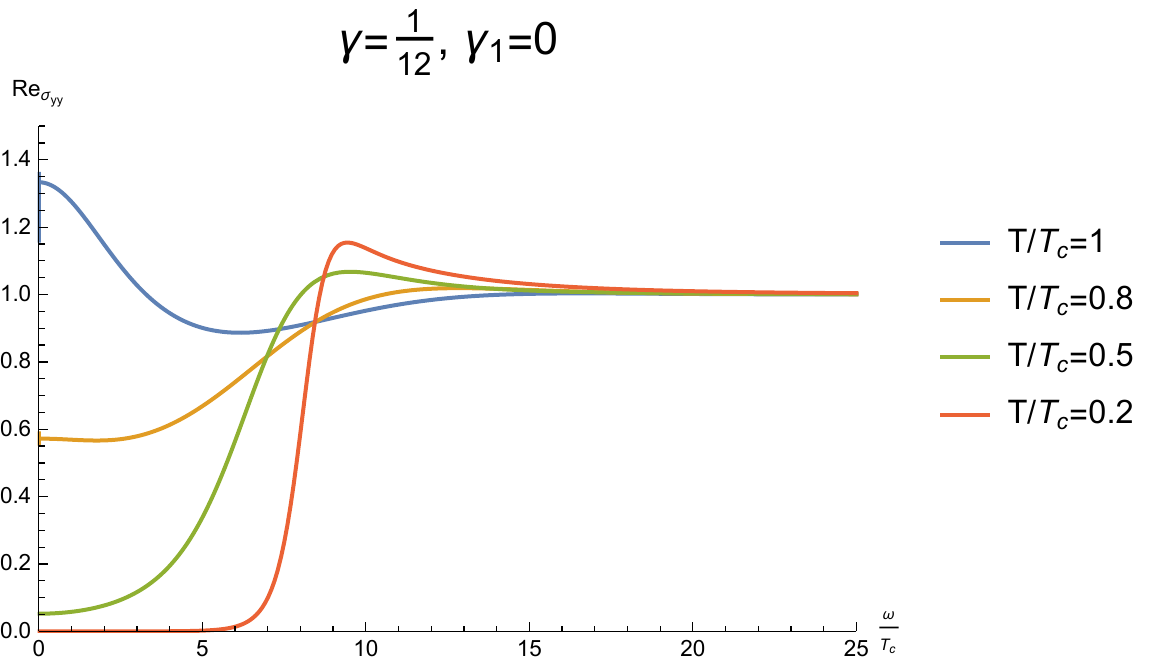}\hspace{0.5cm}
\includegraphics[scale=0.6]{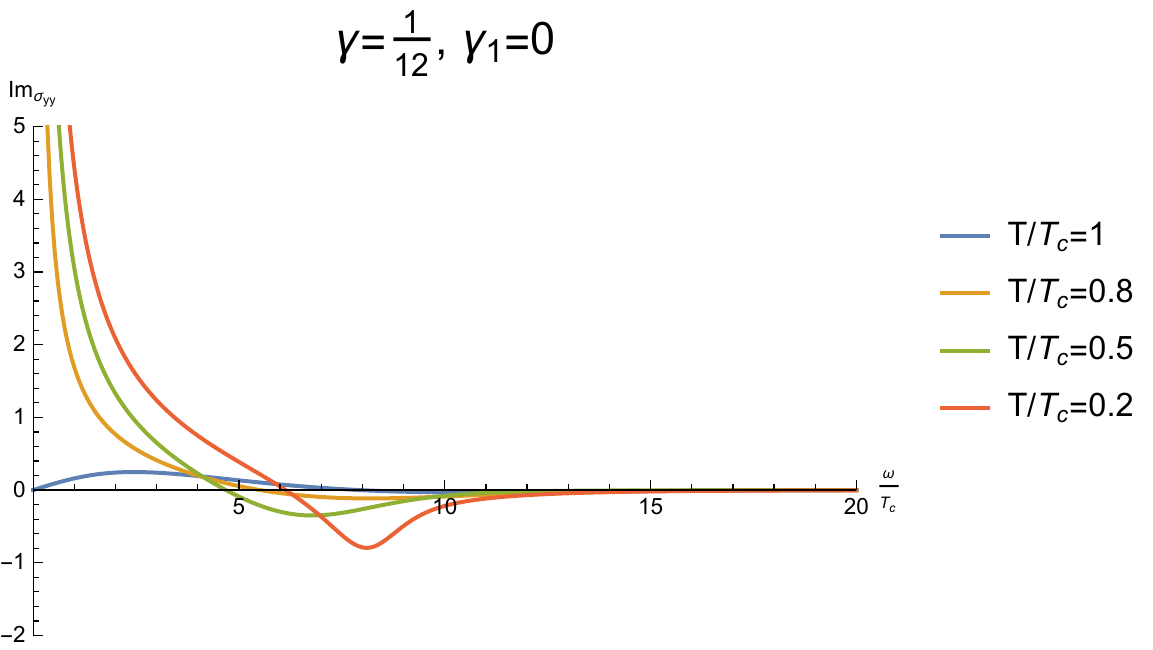} \\
\caption{\label{yyforTgamma} Real and imaginary parts of AC conductivity $\sigma_{yy}$ in $4$ derivative theory.}}
\end{figure*}
\begin{figure*}
\center{
\includegraphics[scale=0.6]{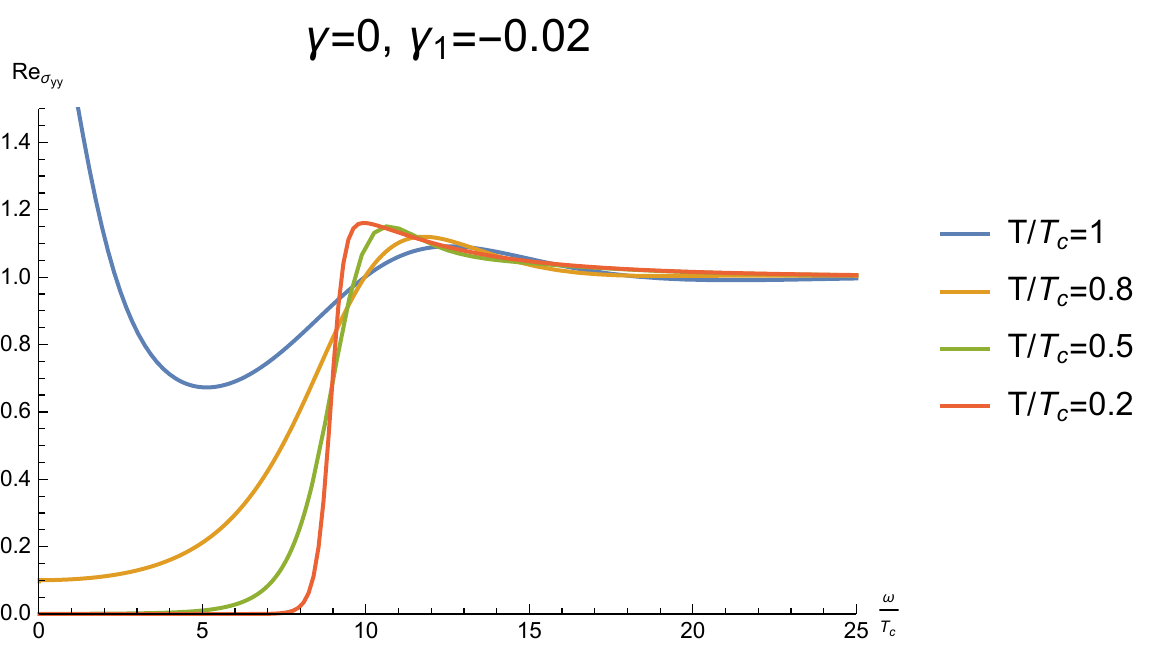}\hspace{0.5cm}
\includegraphics[scale=0.6]{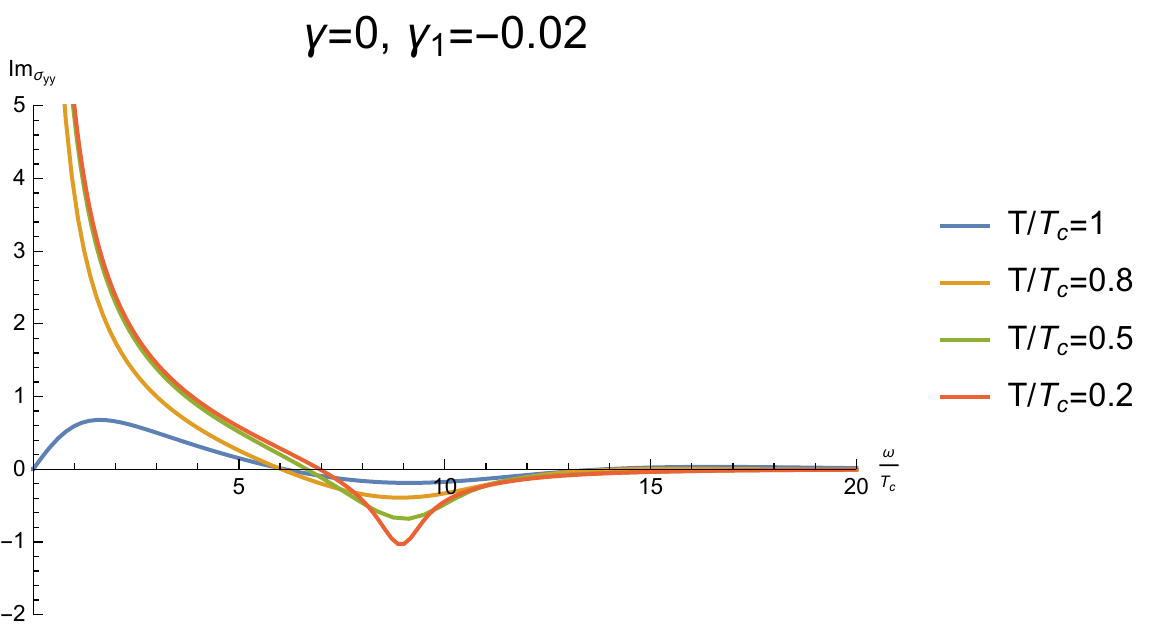} \\
\includegraphics[scale=0.6]{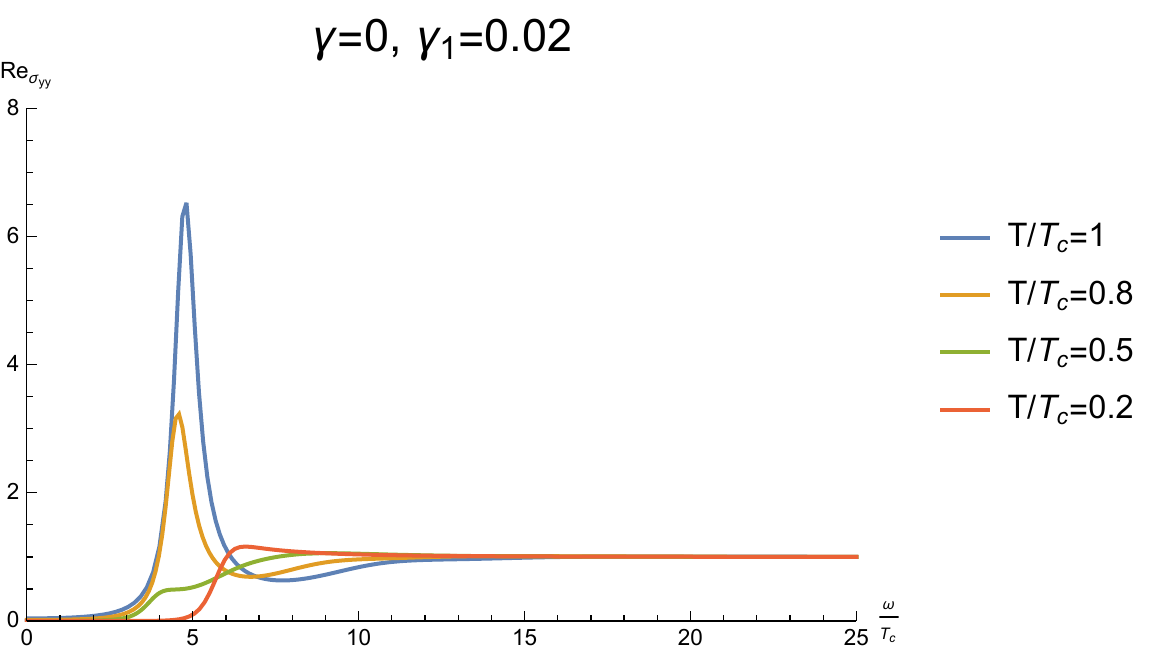}\hspace{0.5cm}
\includegraphics[scale=0.6]{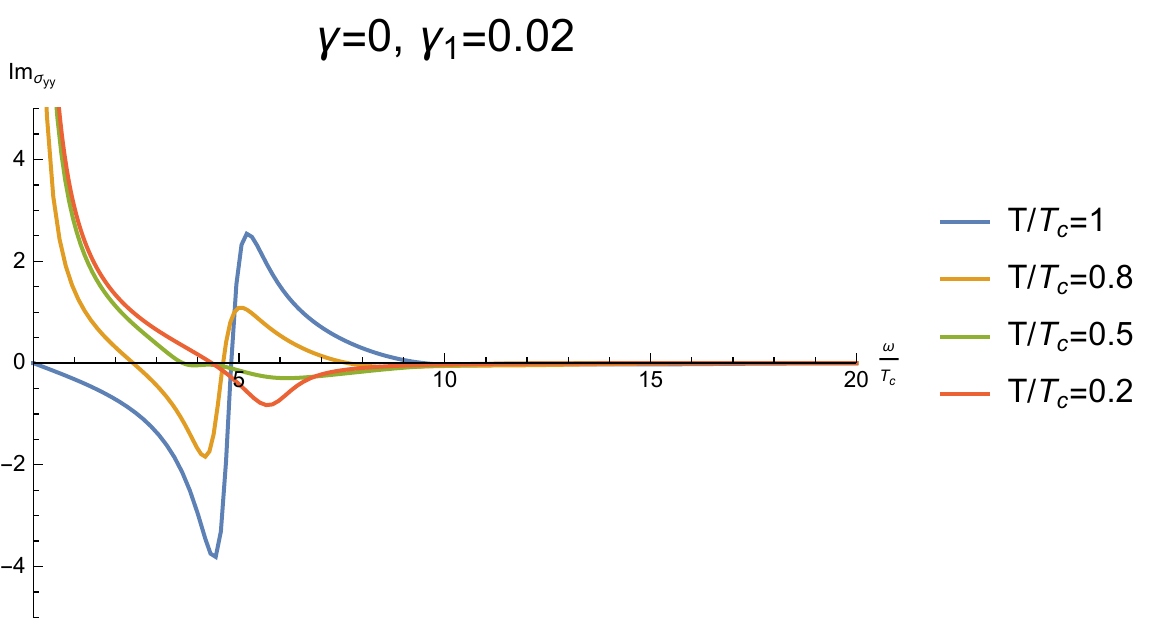} \\
\caption{\label{yyforTgamma1} Real and imaginary parts of AC conductivity $\sigma_{yy}$ in $6$ derivative theory.}}
\end{figure*}

FIG.~\ref{yyforTgamma} and FIG.~\ref{yyforTgamma1} exhibit the real and imaginary parts of AC conductivity $\sigma_{yy}$ for $4$ and $6$ derivative theories, respectively. In particular, we show the evolution of AC conductivity with the temperature from normal state to superconducting state. The main properties of the conductivity are closely similar to that of holographic s-wave superconductor with Weyl corrections studied in \cite{Wu:2010vr,Wu:2017xki}. These properties are summarized as:
\begin{widetext}
\begin{table}[ht]
\begin{center}
\begin{tabular}{|c|c|c|c|c|c|c|c}
 \hline
$~~\gamma~~$ &~~$-1/12$~~&~~$0$~~&~~$1/12$~~
          \\
        \hline
~~$C$~~ & ~~$46.0118$~~ & ~~$31.7119$~~ & ~~$8.9811$~~
          \\
\hline
\end{tabular}
\caption{\label{c-table} The constant $C$ in Eq. \eqref{ns} with different $\gamma$. Here $\gamma_{1}=0$.}
\end{center}
\end{table}
\end{widetext}
\begin{widetext}
\begin{table}[ht]
\begin{center}
\begin{tabular}{|c|c|c|c|c|c|c|c}
\hline
$~~\gamma_{1}~~$&~~$-0.02$~~&~~$0$~~&~~$0.02$~~
          \\
\hline
~~$C$~~ & ~~$38.6926$~~ & ~~$31.7119$~~ & ~~$7.5796$~~
          \\
\hline
\end{tabular}
\caption{\label{c-table1} The constant $C$ in Eq. \eqref{ns} with different $\gamma_{1}$. Here $\gamma=0$.}
\end{center}
\end{table}
\end{widetext}
\begin{itemize}
  \item In the normal state, thanks to the electromagnetic (EM) self-duality, AC conductivity without HD term is independent of the frequency. The introduction of the HD terms breaks the EM self-duality and so leads to the frequency dependent AC conductivity (FIG.~\ref{yyforTgamma} and FIG.~\ref{yyforTgamma1}).
  \item At $\omega=0$, a pole emerges in the imaginary part of the conductivity (right plots in FIG.~\ref{yyforTgamma} and FIG.~\ref{yyforTgamma1}). According to the Kramers-Kronig (KK) relation, there is a corresponding delta function in its real part at $\omega=0$, which suggests the emergence of the superconductivity.
  \item More specifically, in the limit of $\omega=0$, the real part of conductivity follows $Re[\sigma(\omega)]\sim\pi n_{s}\delta(\omega)$ and the imaginary part behaves as $Im[\sigma(\omega)]\sim n_{s}/\omega$, where $n_s$ is the superfluid density. We can fit the data near the critical temperature and find that the superfluid density follows the same behavior for the different couplings $\gamma$ and $\gamma_{1}$ as \cite{Hartnoll:2008vx,Kuang:2016edj}
      \fa
      \label{ns}
      n_{s}\simeq C T_{c}\Big(1-\frac{T}{T_{c}}\Big),
      \ffa
      where $C$ is a constant. We show the values of $C$ for the different couplings $\gamma$ and $\gamma_{1}$ in TABLE~\ref{c-table} and TABLE~\ref{c-table1}, from which we see that the values of $C$ change with the coupling parameters $\gamma$ or $\gamma_{1}$.
      The superfluid density $n_{s}$ vanishes linearly as $T\rightarrow T_{c}$.
      \item
      On the other hand, in the superconducting phase, there is still a contribution of the normal, non-superconducting, component to the DC conductivity, which we refer to as normal fluid density $n_n$ and is defined by
      \fa
      \label{nn}
      n_n=\lim_{\omega\rightarrow 0}Re[\sigma(\omega)]\,.
      \ffa
      We work out the normal fluid density $n_n$ for different temperature in TABLE~\ref{nn-table} and TABLE~\ref{nn-table1}.
      We find that when the system just enters into the superconducting phase, $n_n$ is still finite near the critical temperature. It means that it is coexist of superfluid density $n_s$ and normal fluid density $n_n$ in the superconducting phase.
      Therefore, the system resembles a two-fluid model as the standard holographic superconductor model \cite{Hartnoll:2008vx,Hartnoll:2008kx,Horowitz:2010gk}. As the temperature falls $n_n$ goes down and vanishes at extremal low temperature. It suggests that the normal component of the electron fluid is going down to develop into the superfluid component and vanishes at extremal low temperature.
  \item For $6$ derivative theory with positive $\gamma_1$, the AC conductivity displays a hard-gap-like at low frequency in the normal state. Then a pronounced peak emerges at intermediate frequency \cite{Witczak-Krempa:2013aea}. After the superconducting phase transition happens, the pronounced peak gradually decreases with the temperature (the bottom panels in FIG.~\ref{yyforTgamma1}).
  \item The superconducting energy gap $\omega_g/T_c$ runs with the coupling parameters ranging from approaching the value predicted by BCS theory, to the one of the strong coupling high temperature superconducting energy gap (see FIG.~\ref{yyforTgamma}, TABLE \ref{wg-table} and TABLE \ref{wg-table1}). These observations are in agreement with that of the $U(1)$ gauge field over SS-AdS black brane \cite{Wu:2010vr,Wu:2017xki}.
\end{itemize}
\begin{widetext}
\begin{table}[ht]
\begin{center}
\begin{tabular}{|c|c|c|c|c|c|c|c|}
         \hline
~$n_n$~ &~$\frac{T}{T_{c}}=1$~&~$\frac{T}{T_{c}}=0.95$~&~$\frac{T}{T_{c}}=0.90$~&~$\frac{T}{T_{c}}=0.85$~&~$\frac{T}{T_{c}}=0.80$~&~$\frac{T}{T_{c}}=0.50$~&~$\frac{T}{T_{c}}=0.20$
          \\
        \hline
~$\gamma=-1/12$~& ~$0.5221$ &~$0.2064$~&~$0.0837$~&~$0.0307$& ~$0.0131$~&~$2.0513\times10^{-5}$& ~$5.0903\times10^{-15}$
          \\
        \hline
~$\gamma=0$~ & ~$0.8795$& ~$0.5535$~&~$0.3627$~&~$0.1815$& ~$0.1137$~&~$0.0011$& ~$1.4325\times10^{-10}$
          \\
        \hline
~$\gamma=1/12$~& ~$1.2125$ & ~$1.0802$~&~$0.8418$~&~$0.6937$& ~$0.5571$~&~$0.0522$& ~$5.4112\times10^{-6}$
          \\
        \hline
\end{tabular}
\caption{\label{nn-table} The normal fluid density $n_n$ in the superconducting phase for the different temperature in $4$ derivative theory ($\gamma_{1}=0$).}
\end{center}
\end{table}
\end{widetext}
\subsection{$\sigma_{xx}$}

We study the conductivity along $x$ direction $\sigma_{xx}$ in this subsection. To this end, we need to solve the coupling equations of $A^3_x$, $A^1_t$ and $A^2_t$, which read as
\fa\label{Ax}
&&
{A_{x}^{3}}''+\Big(\frac{f'}{f}+\frac{2u(\gamma+4u^{2}\gamma_{1}f'')(2f''+uf''')}{-3+2u^{2}f''(\gamma+2u^2\gamma_{1}f'')}\Big)
{A_{x}^{3}}'+\frac{\omega^2}{f^2}{A_{x}^{3}}-\frac{\phi\psi}{f^2}{A_{t}^{1}}-\frac{i\omega\psi}{f^2}{A_{t}^{2}}=0\,,
\
\\
&&
\label{At1}
{A_{t}^{1}}''+\frac{4u(\gamma-2u^{2}\gamma_{1}f'')(2f''+uf''')}{3+4u^{2}{f''}(\gamma-u^{2}\gamma_{1}{f''})}{A_{t}^{1}}'
+\frac{\phi\psi}{f}\Big(1+\frac{6u^{2}\gamma f''}{-3-4u^{2}f''(\gamma+u^{2}\gamma_{1}f'')}\Big)A_{x}^{3}=0\,,
\nonumber
\\
&&
\
\\
&&
{A_{t}^{2}}''+\frac{4u(\gamma-2u^{2}\gamma_{1}f'')(2f''+uf''')}{3+4u^{2}{f''}(\gamma-u^{2}\gamma_{1}{f''})}{A_{t}^{2}}'
+\frac{\psi^2}{f}\Big(-1+\frac{6u^2\gamma f''}{3+4u^{2}{f''}(\gamma-u^{2}\gamma_{1}{f''})}\Big){A_{t}^{2}}
\nonumber
\\
&&
\label{At2}
+\frac{i\omega \psi}{f}\Big(-1+\frac{6u^2\gamma f''}{3+4u^{2}{f''}(\gamma-u^{2}\gamma_{1}{f''})}\Big){A_{x}^{3}}=0\,.
\ffa

\begin{widetext}
\begin{table}[ht]
\begin{center}
\begin{tabular}{|c|c|c|c|c|c|c|c|}
         \hline
~$n_n$~ &~$\frac{T}{T_{c}}=1$~&~$\frac{T}{T_{c}}=0.95$~&~$\frac{T}{T_{c}}=0.90$~&~$\frac{T}{T_{c}}=0.85$&~$\frac{T}{T_{c}}=0.80$~&~$\frac{T}{T_{c}}=0.50$&~$\frac{T}{T_{c}}=0.20$
          \\
        \hline
~$\gamma_{1}=-0.02$~& ~$1.7841$ &~$0.8497$~&~$0.4620$~&~$0.2005$& ~$0.0968$~&~$0.0007$& ~$1.8302\times10^{-11}$
          \\
        \hline
~$\gamma_{1}=0$~ & ~$0.8795$& ~$0.5535$~&~$0.3627$~&~$0.1815$& ~$0.1137$~&~$0.0011$& ~$1.4325\times10^{-10}$
          \\
        \hline
~$\gamma_{1}=0.02$~& ~$0.0389$ & ~$0.0323$~&~$0.0267$~&~$0.0191$& ~$0.0156$~&~$0.0014$& ~$1.1349\times10^{-7}$
          \\
        \hline
\end{tabular}
\caption{\label{nn-table1} The normal fluid density $n_n$ in the superconducting phase for the different temperature in $6$ derivative theory ($\gamma=0$).}
\end{center}
\end{table}
\end{widetext}

Notice that there exist two additional constraints:
\fa
&&
\label{constraint1}
 i\omega{A_{t}^{1}}+\phi{A_{t}^{2}}'-{\phi}'{A_{t}^{2}}=0\,,
\
\\
&&
\label{constraint2}
(4u^4\gamma_{1}f''^{2}-4u^2\gamma f''-3)(\phi{A_{t}^{1}}'+i\omega {A_{t}^{2}}')+(4u^4\gamma_{1}f\psi f''^{2}+2u^2\gamma f\psi f''-3f\psi){A_{x}^{3}}'-
\nonumber
\\
&&
(4u^4\gamma_{1}\phi'f''^{2}-4u^2\gamma\phi'f''-3\phi'){A_{t}^{1}}-(4u^4\gamma_{1}f\psi'f''^{2}+2u^2\gamma f\psi'f''-3f\psi')A_{x}^{3}=0\,.
\ffa
The constraints (\ref{constraint1}) and (\ref{constraint2}) are not independent of the EOMs \eqref{Ax}, \eqref{At1} and \eqref{At2}. In fact, the constrained second order equations follow algebraically from the EOMs.

Now, we are ready to numerically solve the EOMs \eqref{Ax}, \eqref{At1} and \eqref{At2}. Near the horizon, one expands $A^3_x$, $A^1_t$ and $A^2_t$ as
\fa
&&
\label{A_{x}}
A_{x}^{3}=(1-u)^{-i\omega/4\pi T}[1+A_{x}^{3(1)}(1-u)+A_{x}^{3(2)}(1-u)^{2}+\cdot \cdot \cdot]\,,
\
\\
&&
\label{A_{t}^{1}}
A_{t}^{1}=(1-u)^{-i\omega/4\pi T}[A_{t}^{1(2)}(1-u)^{2}+A_{t}^{1(3)}(1-u)^{3}+\cdot \cdot \cdot]\,,
\
\\
&&
\label{A_{t}^{2}}
A_{t}^{2}=(1-u)^{-i\omega/4\pi T}[A_{t}^{2(1)}(1-u)+A_{t}^{2(2)}(1-u)^{2}+\cdot \cdot \cdot]\,.
\ffa
where all the coefficients $A_{u}^{a(i)}$ are determined by the model parameters as well as $\omega$. Notice that above we have imposed the ingoing wave conditions at the horizon. Near the UV boundary, the perturbative fields $A^3_x$, $A^1_t$ and $A^2_t$ fall in the following forms
\fa
&&
{A_{x}^{3}}=a_{x}^{3(0)}+u{a_{x}^{3(1)}}+\cdot \cdot \cdot\,,
\
\\
&&
{A_{t}^{1}}=a_{t}^{1(0)}+u{a_{t}^{1(1)}}+\cdot \cdot \cdot\,,
\
\\
&&
{A_{t}^{2}}=a_{t}^{2(0)}+u{a_{t}^{2(1)}}+\cdot \cdot \cdot\,.
\ffa
Through the holographic dictionary, the conductivity $\sigma_{xx}$ can be compute as
\fa
\sigma_{xx}=-\frac{i}{\omega {A_{x}^{3(0)}}}\Big({A_{x}^{3(1)}}+\psi^{(1)}\frac{i\omega{A_{t}^{2(0)}}+\mu{A_{t}^{1(0)}}}{\mu^2-\omega^2}\Big)\,.
\ffa

\begin{widetext}
\begin{table}[ht]
\begin{center}
\begin{tabular}{|c|c|c|c|c|c|c|c}
 \hline
$~~\gamma~~$ &~~$-1/12$~~&~~$0$~~&~~$1/12$~~
          \\
        \hline
~~$\omega_{g}/T_{c}$~~ & ~~$9.3515$~~ & ~~$8.1634$~~ & ~~$6.3019$~~
          \\
\hline
\end{tabular}
\caption{\label{wg-table} The superconducting energy gap $\omega_{g}/T_{c}$ of $\sigma_{yy}$ with different $\gamma$. Here $\gamma_{1}=0$.}
\end{center}
\end{table}
\end{widetext}
\begin{widetext}
\begin{table}[ht]
\begin{center}
\begin{tabular}{|c|c|c|c|c|c|c|c}
\hline
$~~\gamma_{1}~~$&~~$-0.02$~~&~~$0$~~&~~$0.02$~~
          \\
\hline
~~$\omega_{g}/T_{c}$~~ & ~~$8.9350$~~ & ~~$8.1634$~~ & ~~$5.7065$~~
          \\
\hline
\end{tabular}
\caption{\label{wg-table1} The superconducting energy gap $\omega_{g}/T_{c}$ of $\sigma_{yy}$ with different $\gamma_{1}$. Here $\gamma=0$.}
\end{center}
\end{table}
\end{widetext}

\begin{figure*}
\center{
\includegraphics[scale=0.65]{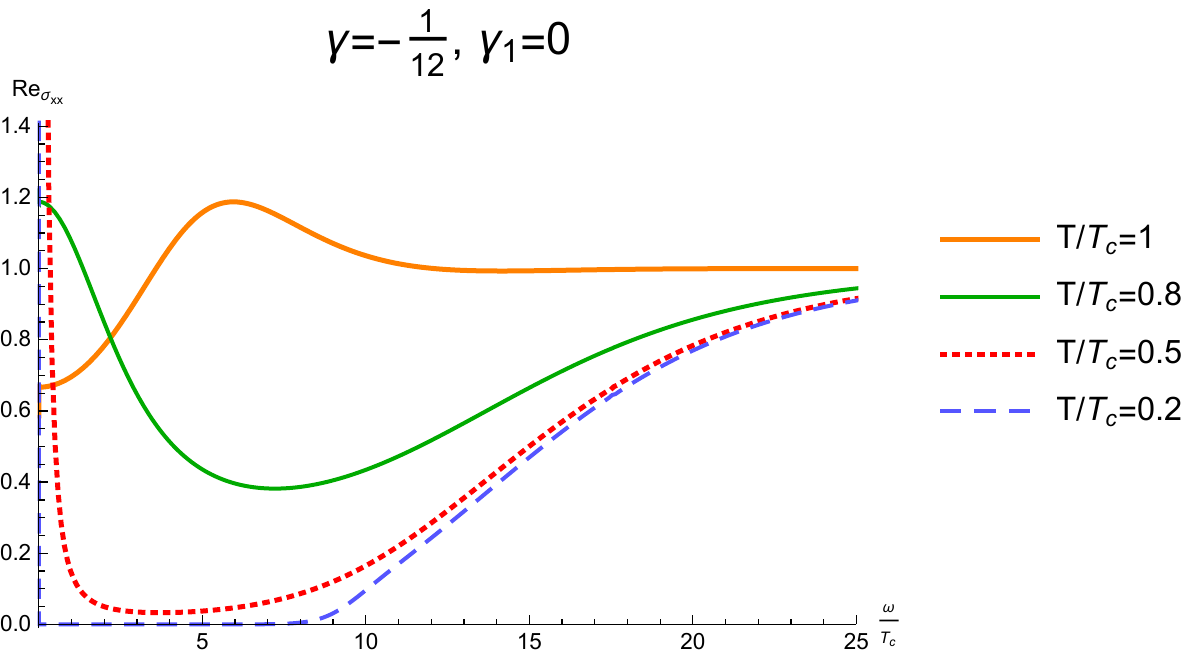}\hspace{0.1cm}
\includegraphics[scale=0.65]{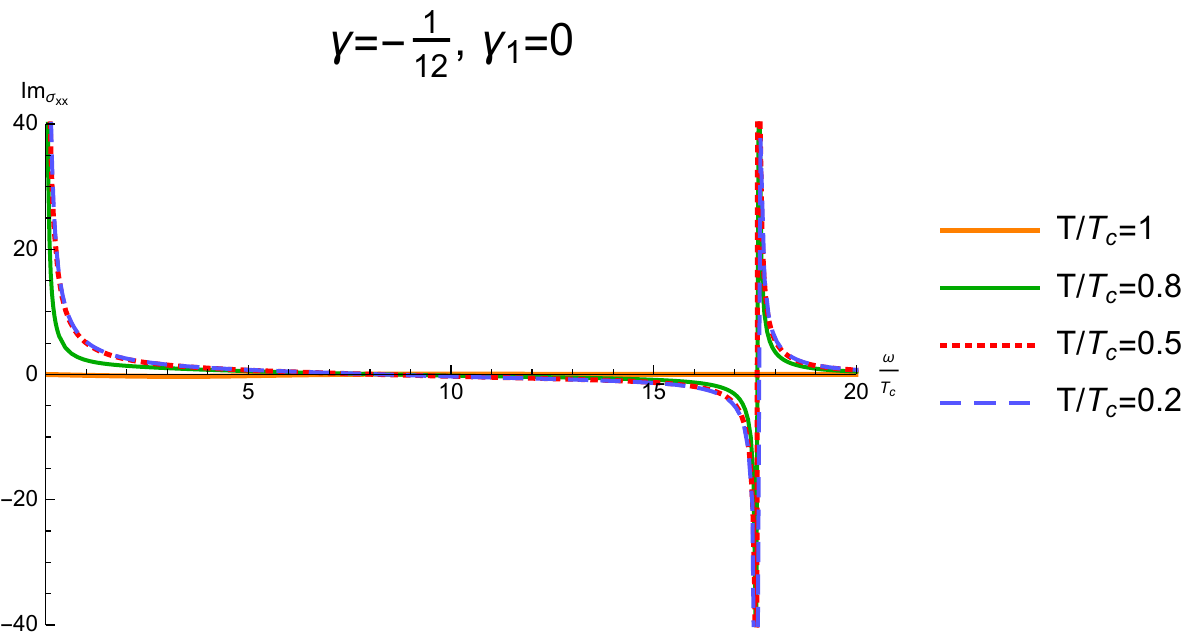} \\
\includegraphics[scale=0.65]{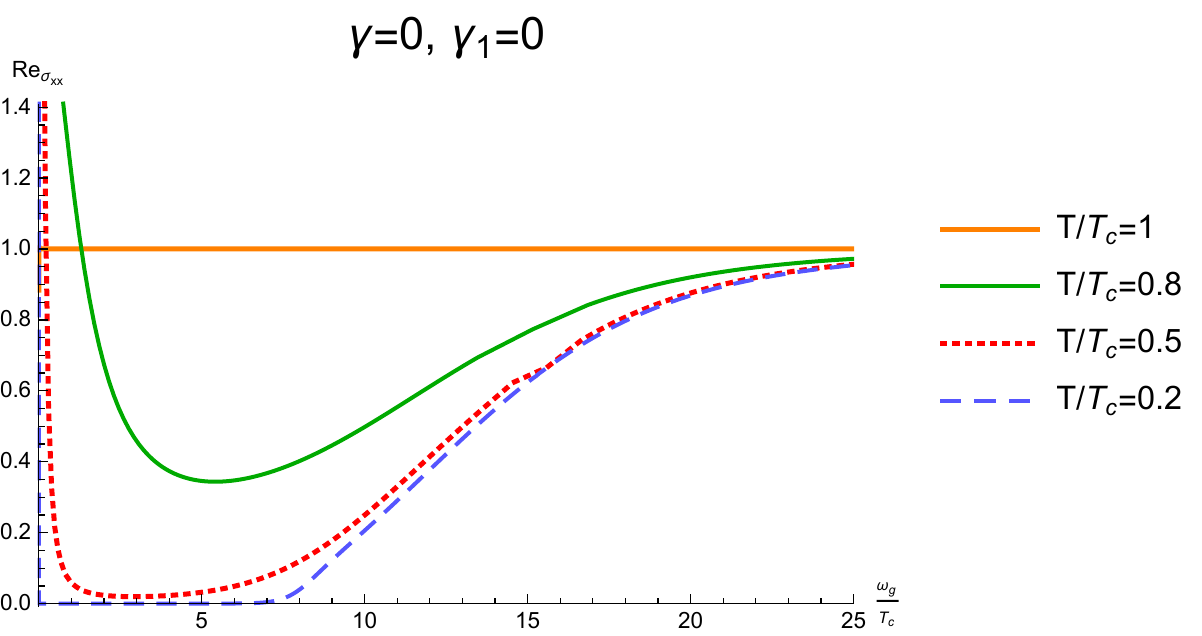}\hspace{0.1cm}
\includegraphics[scale=0.65]{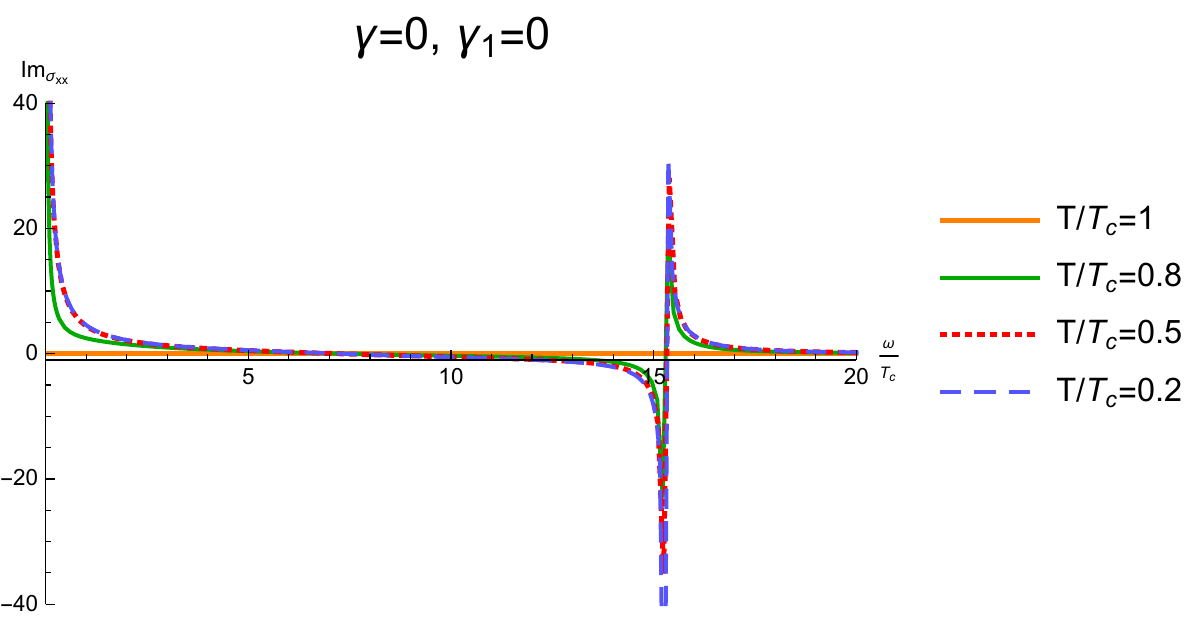} \\
\includegraphics[scale=0.65]{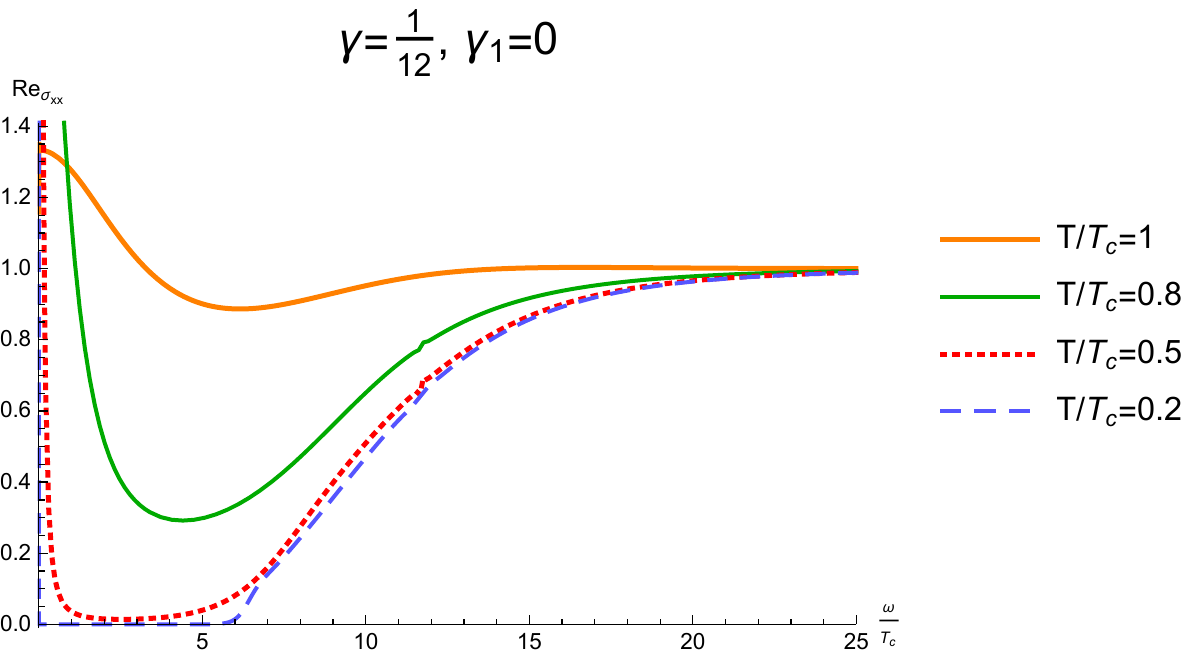}\hspace{0.1cm}
\includegraphics[scale=0.65]{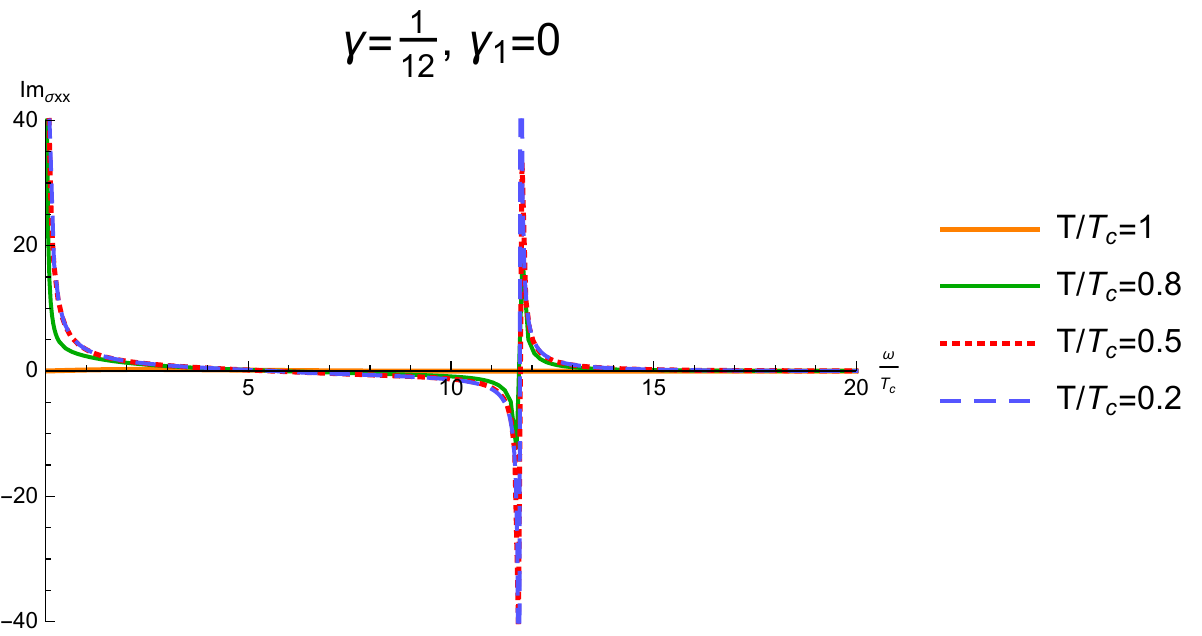} \\
\caption{\label{xxv1} The conductivities $\sigma_{xx}$ as a function of the frequency for different $\gamma$. The left panels are for the real part and the right panels are for the imaginary one.}}
\end{figure*}
\begin{figure*}
\center{
\includegraphics[scale=0.65]{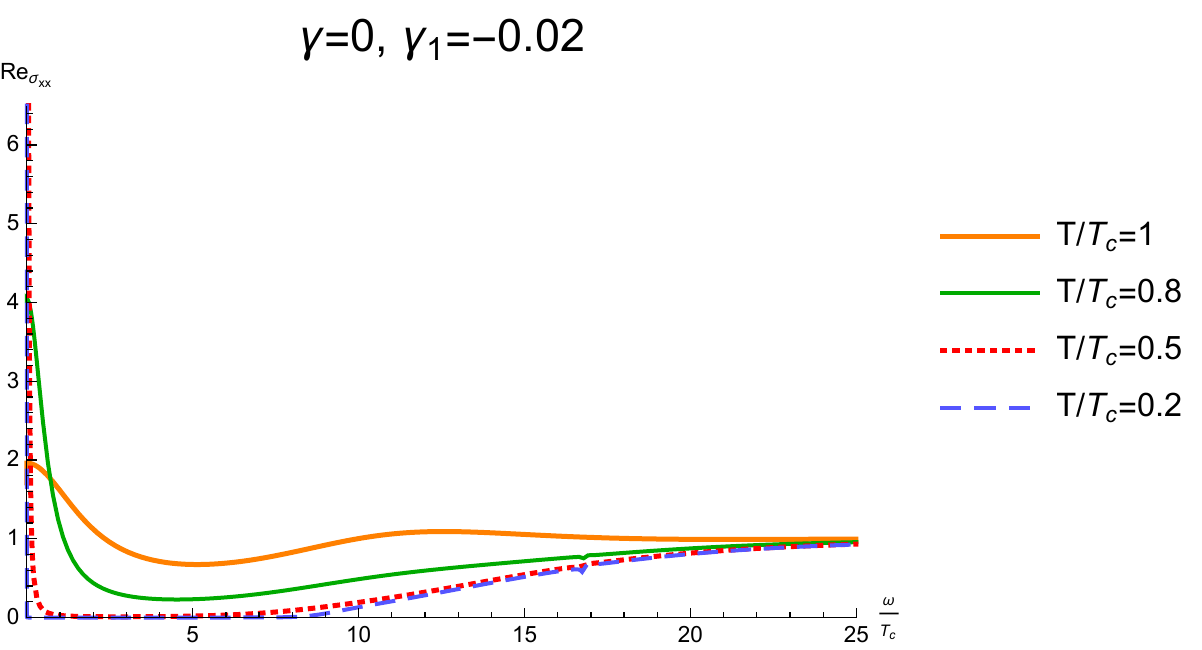}\hspace{0.1cm}
\includegraphics[scale=0.65]{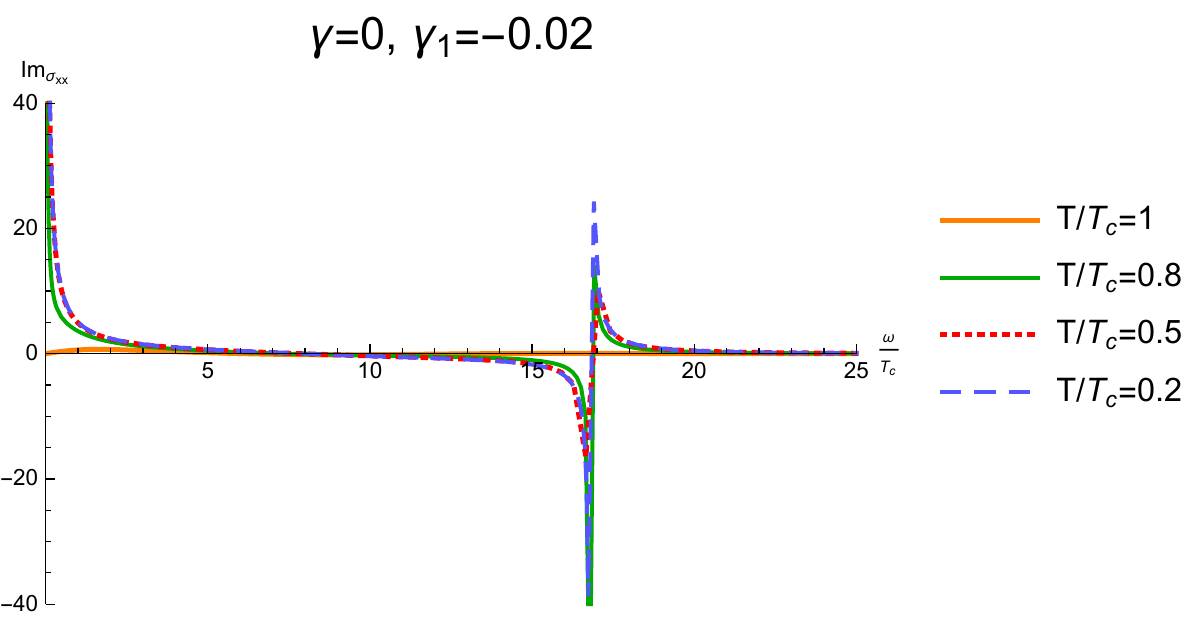} \\
\includegraphics[scale=0.65]{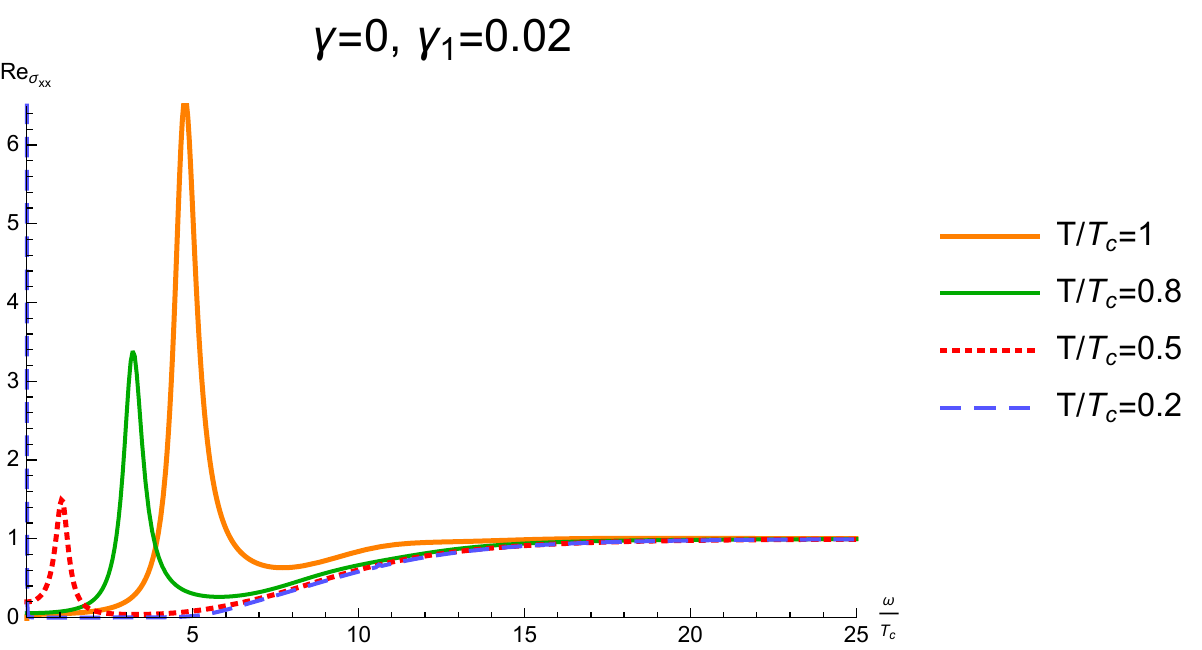}\hspace{0.1cm}
\includegraphics[scale=0.65]{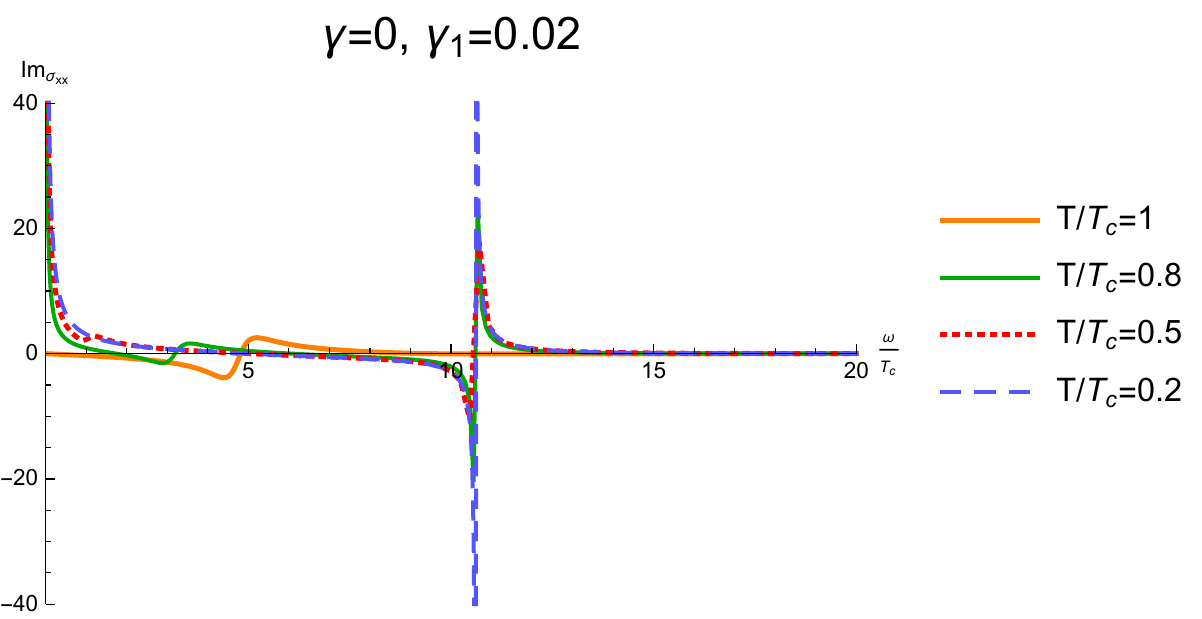} \\
\caption{\label{xxv2} The conductivities $\sigma_{xx}$ as a function of the frequency for different $\gamma_1$. The left panels are for the real part and the right panels are for the imaginary one.}}
\end{figure*}

We show the numerical results in FIG.~\ref{xxv1} and FIG.~\ref{xxv2}. The behaviors of the real part of $\sigma_{xx}$ in the normal state are closely similar to that of $\sigma_{yy}$. That is to say, a dip exhibits in low frequency for $\gamma=-1/12$, while a peak emerges for $\gamma=1/12$ in $4$ derivative theory. In $6$ derivative theory, a Drude-like peak exhibits in low frequency for $\gamma_1=-0.02$, while for $\gamma_1=0.02$, we observe a hard-gap-like behavior at low frequency and a pronounced peak at intermediate frequency. Therefore, in the normal state, we cannot observe an obvious anisotropic behavior.

However, when the systems enter into the superconducting phase, the obvious anisotropic behavior can be observed as previous studies \cite{Gubser:2008wv,Kuang:2011dy,Cai:2010cv}. The most important feature of $\sigma_{xx}$ is that a Drude-like peak in $Re[\sigma_{xx}]$ shows up in the frequency. Especially, this feature is independent of the model parameters. In addition, with the decrease of the temperature, the Drude-like behavior becomes more evident. In $Im[\sigma_{xx}]$, a pole in high frequency emerges. It means that the anisotropic effect plays a dominant role in the behavior of $\sigma_{xx}$.
\begin{figure*}
\center{
\includegraphics[scale=0.55]{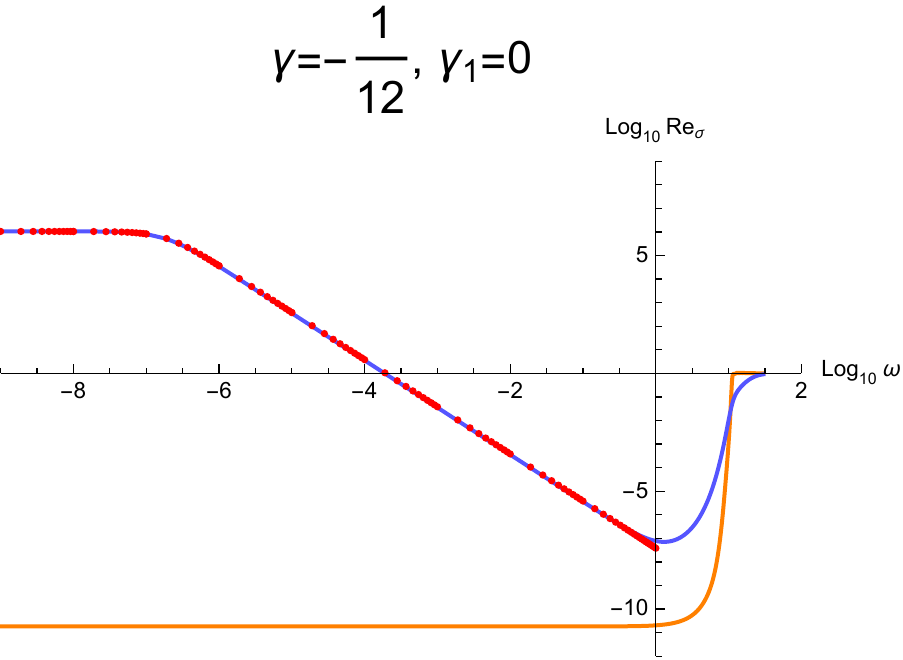}\hspace{0.3cm}
\includegraphics[scale=0.55]{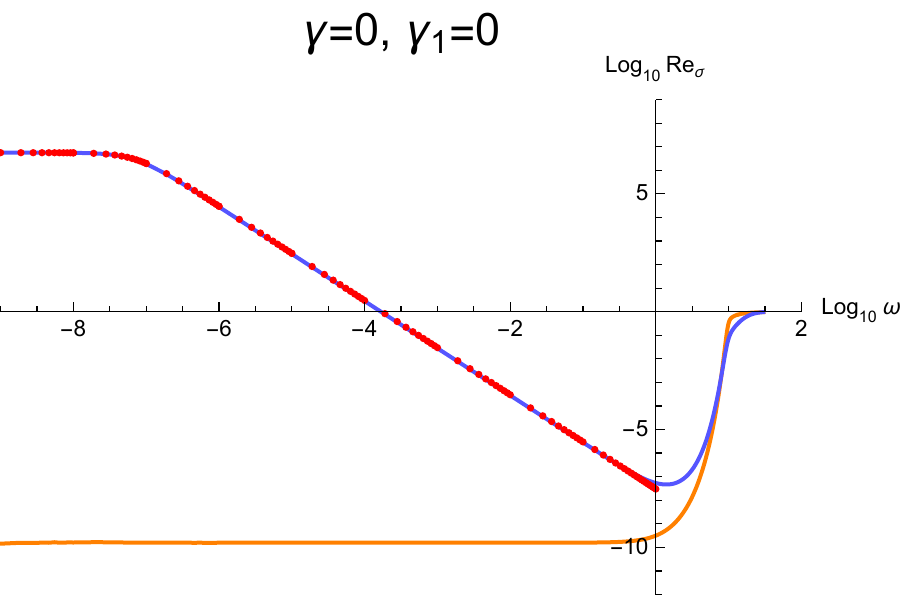}\hspace{0.3cm}
\includegraphics[scale=0.55]{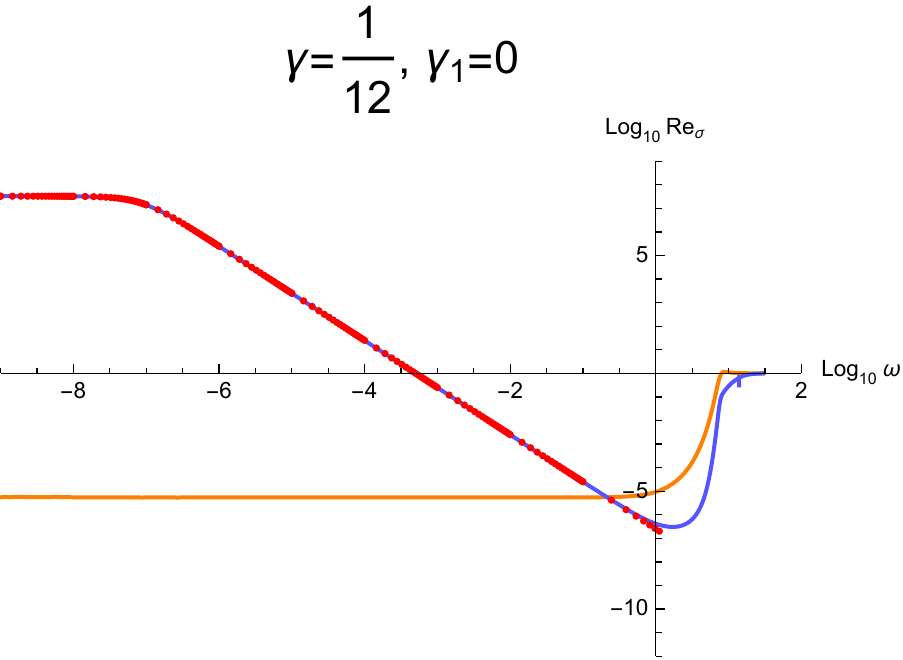}\hspace{0.3cm}
\caption{\label{Fitgamma} Logarithmic real parts of conductivity versus the logarithmic frequencies in $4$ derivative theory with different coupling parameter $\gamma$. The blue line is $\sigma_{xx}$ while the orange line for $\sigma_{yy}$. The red points are the fitting of $\sigma_{xx}$ for various $\gamma$ in the low frequency limit.}}
\end{figure*}
\begin{figure*}
\center{
\includegraphics[scale=0.55]{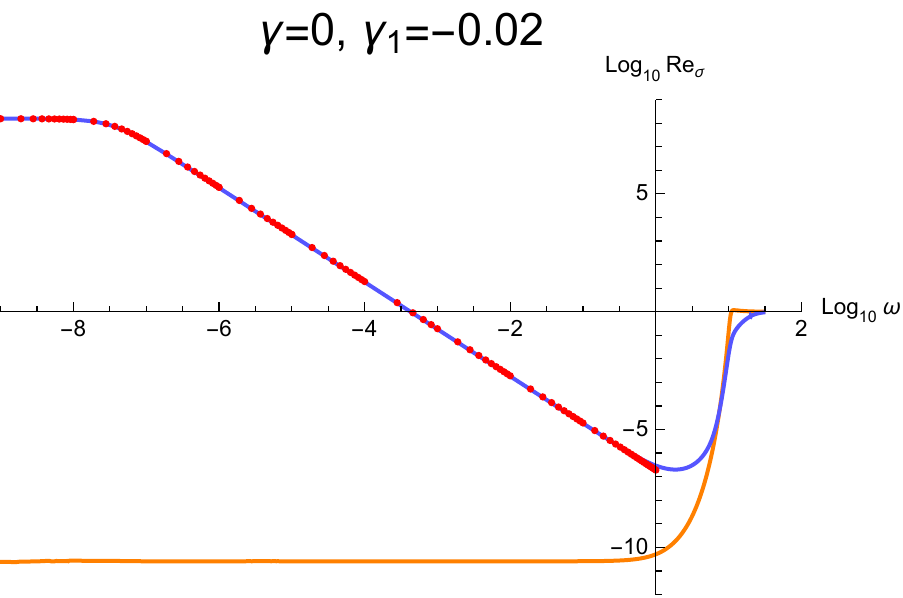}\hspace{0.3cm}
\includegraphics[scale=0.55]{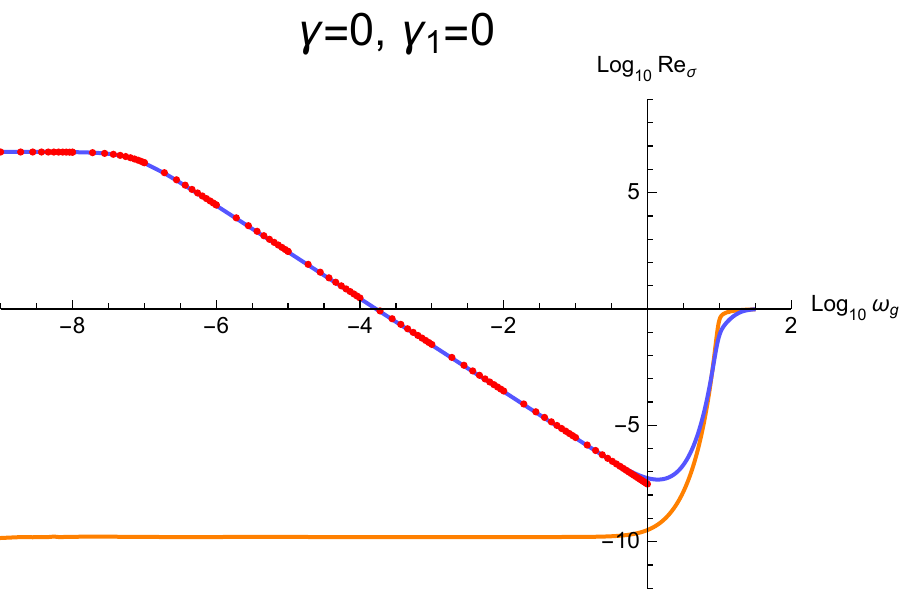}\hspace{0.3cm}
\includegraphics[scale=0.55]{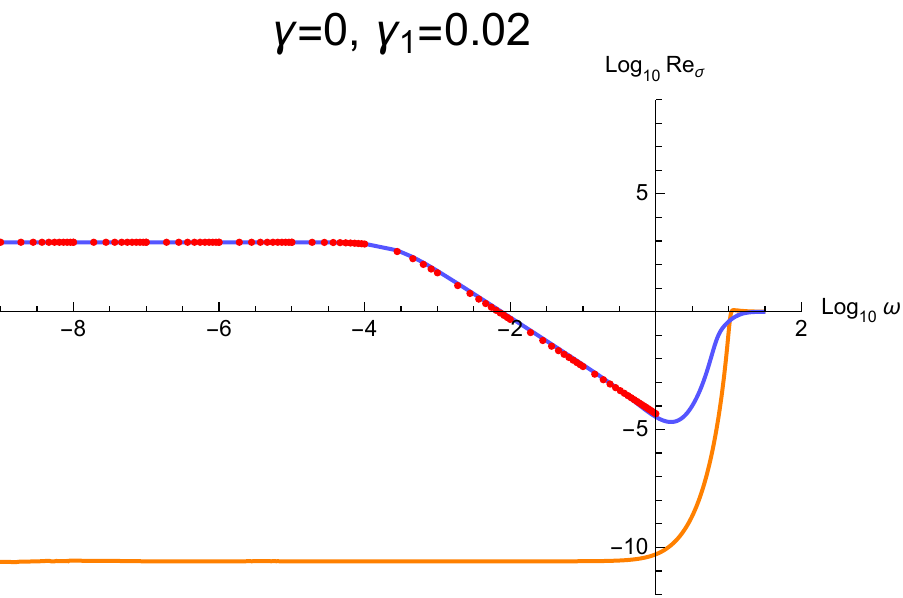}\hspace{0.3cm}
\caption{\label{Fitgamma1} Logarithmic real parts of conductivity versus the logarithmic frequencies in $6$ derivative theory with different coupling parameter $\gamma_{1}$. The blue line is $\sigma_{xx}$ while the orange line for $\sigma_{yy}$. The red points are the fitting of $\sigma_{xx}$ for various $\gamma_{1}$ in the low frequency limit.}}
\end{figure*}

Finally, we make a special focus on the Drude-like behavior of $\sigma_{xx}$, which is firstly noticed in~\cite{Gubser:2008wv} and further studied in~\cite{Kuang:2011dy,Cai:2010cv}. The real part of the conductivity for Drude model follows the following behavior
\fa\label{Drude}
\rm Re(\sigma)_{\rm Durde}=\frac{\sigma_{0}}{1+\omega^{2}\tau^{2}}\,,
\ffa
where $\sigma_{0}=ne^{2}\tau/m$ is a constant related to the electron density, charge, mass and the relaxation time. Based on the formula~(\ref{Drude}), we can fit our numerical results to evaluate $\sigma_{0}$ and $\tau$ in low frequencies and the fitting results of $\sigma_{0}$ and $\tau$ are shown in the TABLE~\ref{Fittinggamma} and TABLE~\ref{Fittinggamma1}. In addition, we plot the relations between $\log_{10}[Re]$ and $\log_{10}[\omega]$ for fixed different $\gamma$ and $\gamma_{1}$ in Fig~\ref{Fitgamma} and Fig~\ref{Fitgamma1}. It is easily find that the DC conductivity $\sigma_{0}$ and the relaxation time $\tau$ depend on the coupling parameters. The $\sigma_{0}$ increases with the $\gamma$ growing and the $\tau$ increases first and then goes down with the $\gamma$ growing, while for the case of $\gamma_{1}$, the monotonous decreasing changing tendencies of $\sigma_{0}$ and $\tau$ was observed. From FIG.~\ref{Fitgamma} and FIG.~\ref{Fitgamma1}, the anisotropic behavior of conductivity is very remarkable.

\begin{widetext}
\begin{table}[ht]
\begin{center}
\begin{tabular}{|c|c|c|c|c|c|}
\hline $\gamma$ &$\log_{10}[\sigma_0]$ &$\log_{10}[\tau]$\\
\hline -1/12 &$6.0216$ & $5.2517\times10^6$\\
\hline 0 & $6.7471$ &$ 1.3677\times10^7$\\
\hline 1/12 & $7.5161$ & $1.1245\times10^7$ \\
\hline
\end{tabular}
\caption{\label{Fittinggamma} The values of $\log_{10}[\sigma_0]$ and $\log_{10}[\tau]$ from the fitting points in $4$ derivative theory with different $\gamma$. Here $\gamma_{1}=0$.}
\end{center}
\end{table}
\end{widetext}
\begin{widetext}
\begin{table}[ht]
\begin{center}
\begin{tabular}{|c|c|c|c|c|c|}
 \hline $\gamma_{1}$ &$\log_{10}[\sigma_0]$ &$\log_{10}[\tau]$\\
\hline -0.02 &$8.1874$ & $2.8467\times10^7$\\
\hline 0 & $6.7471$ &$ 1.3677\times10^7$\\
\hline 0.02 & $2.9457$ & $4.2623\times10^3$ \\
 \hline
 \end{tabular}
\caption{\label{Fittinggamma1} The values of $\log_{10}[\sigma_0]$ and $\log_{10}[\tau]$ from the fitting points in $6$ derivative theory with different $\gamma$. Here $\gamma=0$.}
\end{center}
\end{table}
\end{widetext}

\section{Conclusions and discussions}\label{sec-Conclusions}

In this paper, we construct a holographic $p$-wave superconductor model in a four-dimensional bulk spacetimes with Weyl corrections including the $4$ derivative term and the $6$ derivative term. The numerical results indicate that the superconducting phase happens with the decrease of the temperature and the HD terms do not seem to spoil the generation of the p-wave superconducting phase. And then, we mainly study the properties of AC conductivity. For p-wave superconductor, the conductivity exhibits anisotropic behaviors. The features of the conductivity $\sigma_{yy}$ are very closely similar to that of holographic s-wave superconductor. The main characteristics are summarized as what follows.
\begin{itemize}
  \item From the behaviors of DC conductivity along $y$ direction, i.e., coexistence of superfluid density $n_s$ and normal fluid density $n_n$ in the superconducting phase, we can conclude that the systems are two-fluid models.
  \item But for $6$ derivative theory with positive $\gamma_1$, the AC conductivity displays a hard-gap-like at low frequency in the normal state and so the DC conductivity vanishes. When the system enters into the superconducting phase, the pronounced peak at intermediate frequency is gradually decreasing to form the superfluid component. Therefore, the formation of the superfluid component for $6$ derivative theory with positive $\gamma_1$ is different from that for $4$ derivative theory and $6$ derivative theory with negative $\gamma_1$.
  \item The superconducting energy gap can be observed in the conductivity $\sigma_{yy}$, which is similar to that of holographic s-wave superconductor. The superconducting energy gap runs with the coupling parameters.
\end{itemize}

The behaviors of the real part of the conductivity $\sigma_{xx}$ in the normal state are very similar to that of $\sigma_{yy}$. However, the anisotropy of the conductivity obviously shows up in the superconducting phase. A Drude-like peak at low frequency emerges in $Re[\sigma_{xx}]$ once the system enter into the superconducting phase, regardless of Drude-like peak or hard-gap-like in normal state. Therefore, anisotropy plays a dominant role in the behavior of $\sigma_{xx}$.

\begin{acknowledgments}
We are very grateful to Xiao-Mei Kuang and Peng Liu for helpful discussions and suggestions.
This work is supported by the Natural Science Foundation of China under
Grant Nos. 11775036, 11975072, and 11690021, and Fok Ying Tung Education Foundation under Grant No. 171006.
Guoyang Fu is supported by the Postgraduate Research \& Practice Innovation Program of Jiangsu Province (KYCX20\_2973).
J. P. Wu is also supported by Top Talent Support Program from Yangzhou University. X. Zhang is also supported by the Liaoning Revitalization Talents Program under Grant No. XLYC1905011.

\end{acknowledgments}


\end{document}